\documentclass[manuscript]{aastex}

\newcommand{\HeI}{\ion{He}{1}}

\newcommand{\NeIII}{\ion{Ne}{3}}
\newcommand{\OIII}{\ion{O}{3}}

\newcommand{\GALEX}{{\it GALEX}}

\newcommand{\HST}{{\it HST}}
\newcommand{\IUE}{{\it IUE}}

\newcommand{\Teff}{T_{\rm eff}}

\newcommand{\Spitzer}{{\it Spitzer}}

\newcommand{\WISE}{{\it WISE}}

\begin{document}


\title{The Nucleus of the Planetary Nebula EGB~6 as a Post-Mira
Binary\altaffilmark{1}}

\altaffiltext{1} {Based in part on data obtained with the NASA/ESA {\it Hubble
Space Telescope}, obtained by the Space Telescope Science Institute. STScI is
operated by the Association of Universities for Research in Astronomy, Inc.,
under NASA contract NAS5-26555. Also based in part on observations with the
1.5-m telescope operated by the SMARTS Consortium at Cerro Tololo Inter-American
Observatory.}

\author{Howard E. Bond\altaffilmark{2,3,4}, 
Robin Ciardullo\altaffilmark{2},
Taran L. Esplin\altaffilmark{2},
Steven A. Hawley\altaffilmark{5}, 
James Liebert\altaffilmark{6}, 
and
Ulisse Munari\altaffilmark{7}}

\altaffiltext{2}{Department of Astronomy \& Astrophysics, Pennsylvania State
University, University Park, PA 16802; heb11@psu.edu}

\altaffiltext{3}{Space Telescope Science Institute, 3700 San Martin Drive,
Baltimore, MD 21218}

\altaffiltext{4}
{Visiting astronomer, Kitt Peak National Observatory and Cerro Tololo
Inter-American Observatory, National Optical Astronomy Observatory, which are
operated by the Association of Universities for Research in Astronomy (AURA)
under a cooperative agreement with the National Science Foundation.  
}

\altaffiltext{5}{Department of Physics \& Astronomy, University of Kansas,
Lawrence, KS 66045}

\altaffiltext{6}{Steward Observatory, University of Arizona, Tucson AZ 85721}

\altaffiltext{7}{INAF Astronomical Observatory of Padova, via
dell'Osservatorio 8, 36012 Asiago (VI), Italy}

\begin{abstract}

EGB~6 is a faint, large, ancient planetary nebula (PN)\null. Its central star, a
hot DAOZ white dwarf (WD), is a prototype of a rare class of PN nuclei
associated with dense, compact emission-line knots. The central star also shows
excess fluxes in both the near- (NIR) and mid-infrared (MIR). In a 2013 paper,
we used {\it Hubble Space Telescope\/} (\HST\/) images to show that the compact
nebula is a point-like source, located $0\farcs16$ ($\sim$118~AU) from the
WD\null. We attributed the NIR excess to an M dwarf companion star, which
appeared to coincide with the dense emission knot. We now present new
ground-based NIR spectroscopy, showing that the companion is actually a much
cooler source with a continuous spectrum, apparently a dust-enshrouded
low-luminosity star. New \HST\/ images confirm common proper motion of the
emission knot and red source with the WD\null. The $I$-band, NIR, and MIR fluxes
are variable, possibly on timescales as short as days. We can fit the
spectral-energy distribution with four blackbodies (the WD, a $\sim$1850~K NIR
component, and MIR dust at 385 and 175~K). Alternatively, we show that the
NIR/MIR SED is very similar to that of Class 0/I young stellar objects. We
suggest a scenario in which the EGB~6 nucleus is descended from a wide binary
similar to the Mira system, in which a portion of the wind from an AGB star was
captured into an accretion disk around a companion star; a remnant of this disk
has survived to the present time, and is surrounded by gas photoionized by UV
radiation from the WD.

\end{abstract}

\keywords{white dwarfs --- planetary nebulae --- binaries: visual --- stars:
individual (PG~0950+139) --- planetary nebulae: individual (EGB~6)} 

\clearpage

%

\section{The Case-Book of EGB~6}

EGB~6 (PN~G221.5+46.3) is a low-surface-brightness, high-Galactic-latitude, 
angularly large ($13'\times11'$) planetary nebula (PN), discovered by one of us
in 1978---purely by chance---while examining Palomar Observatory Sky Survey
(POSS) prints. The PN appears close to spherical, but its southwestern edge has
a bright rim, possibly indicating an interaction with the interstellar
medium\footnote{Deep CCD images of EGB~6 have been published by Jacoby \& van de
Steene (1995) and Tweedy \& Kwitter (1996). An excellent deep color image has
been posted on his website by amateur astronomer Dr.\ Don Goldman: {\tt
http://astrodonimaging.com\slash gallery\slash
egb-6-faint-planetary-nebula-in-leo} .}. The object was included in a list of
faint nebulae found on POSS prints by Ellis et al.\ (1984, hereafter EGB)\null.
EGB noted that the POSS photographs show a blue 16th-mag star near the center of
the nebula. This star had also been detected independently by Green et al.\
(1986) during the Palomar-Green survey for high-latitude blue objects, and was
designated PG~0950+139. Follow-up spectroscopy by EGB revealed that this
candidate planetary-nebula nucleus (PNN) has strong [\ion{O}{3}] emission lines,
making it elementary that it is the central star. Subsequently, Fleming et al.\
(1986) classified the absorption-line spectrum of the PNN as that of a hot DA
white dwarf (WD)\null. However, Liebert et al.\ (1989, hereafter L89) showed
that \ion{He}{2} is seen in absorption in the optical spectrum, and later
Gianninas et al.\ (2010, hereafter G10) found absorption lines of heavy elements
in the far ultraviolet. Thus the spectral type of the central star is DAOZ\null.
Based on a non-LTE model-atmosphere analysis, G10 derived stellar parameters of
$\Teff=93,230$~K and $\log g = 7.36$.

The [\ion{O}{3}] emission from the PNN is too strong to come from the
surrounding large, faint, ancient PN\footnote{However, very weak [\OIII]
5007~\AA\ emission from the large PN has been detected serendipitously in the
SDSS spectra of two faint galaxies that happen to lie behind EGB~6: Yuan \& Liu
(2013). Acker et al.\ (1992) list the relative intensities of H$\alpha$,
H$\beta$, and 5007~\AA.}\null. Observations presented by L89 showed that the
nebular lines arise from a compact emission knot (CEK), which is unresolved and
appears to coincide with the PNN in ground-based images. Moreover, the electron
density of the CEK is remarkably high, about $2.2\times10^6\,\rm cm^{-3}$,
according to an emission-line analysis by Dopita \& Liebert (1989, hereafter
DL89).

In addition to the hot WD and associated CEK, there is a compact near-infrared
(NIR) source located near the EGB~6 nucleus. This was first revealed through NIR
photometry of the star by Zuckerman et al.\ (1991), and confirmed by Fulbright
\& Liebert (1993, hereafter FL93). These studies showed that the $J,$ $H$, and
$K$ fluxes exceed those expected from the hot PNN\null. Both sets of authors
concluded that the NIR excess is consistent with the presence of an M dwarf
companion star. Subsequent photometry presented by De~Marco et al.\ (2013) and
Douchin et al.\ (2015, hereafter D15) found no significant excess at the $I$
band, but based on the absolute magnitude of the excess at SDSS $z$ and at $J$,
and an adopted distance, they inferred that the NIR source is a star of spectral
type M3-5~V\null.  However, there had been no direct spectroscopic confirmation
that an M dwarf is present, and it should be remembered ``how dangerous it
always is to reason from insufficient data'' (Conan Doyle 1892)\footnote{In
homage to the canon of Sir Arthur Conan Doyle's fictional character Sherlock
Holmes, a master of scientific investigation, we have included a number of
allusions to these works throughout the text.}. Indeed, De~Marco et al.\ (2013)
noted the curious incident that the $J-H$ color is redder than expected for an
M3-5 spectral type, and Miszalski et al.\ (2011) pointed out that the NIR source
was too luminous for an unreddened star of the observed color.  


%

Yet another component of the system was discovered in {\it Spitzer Space
Telescope\/} observations by Chu et al.\ (2011, hereafter C11), who showed that
the nucleus of EGB~6 has a strong mid-infrared (MIR) excess. C11 surveyed 71 hot
WDs (including 35 that are PNNi) with the \Spitzer\/ multi-band imaging
photometer (MIPS) at $24\,\mu$m, detecting nine of them at fluxes two or more
orders of magnitude above those expected from the hot WDs. Follow-up
observations were obtained by C11 with the \Spitzer\/ infrared array camera
(IRAC) in its four bands from 3.6 to $8.0\,\mu$m. Of the nine $24\,\mu$m
sources, EGB 6 is the apparently brightest of all in the four IRAC bands, and
second brightest at $24\,\mu$m.  C11 modeled its MIR excess as arising from two
cool dust shells, with blackbody temperatures of about 500~K and 150~K\null. 

An extensive discussion of the astrophysical mysteries of the EGB~6 central
system, with additional literature references, was published recently by Liebert
et al.\ (2013, hereafter L13). L13 included results from broad-band and
emission-line imaging and grism spectroscopy with the {\it Hubble Space
Telescope\/} (\HST) obtained between 1991 and 1995, but which had not been
presented in detail previously. The \HST\/ images and spectra revealed the
startling result that the CEK is not centered on the PNN, but is instead a
separate object with a stellar profile, lying $0\farcs166$ away from the hot
nucleus. This corresponds to a projected linear separation of $\sim$118~AU, for
a nominal distance of about 725~pc. The longest-wavelength broad-band \HST\/
images available to L13 were in the F785LP filter of the original Wide
Field\slash Planetary Camera (WF/PC1) and the F814W filter of the Wide Field
Planetary Camera~2 (WFPC2)---that is, filters covering roughly the ground-based
$I$ band. These images revealed a faint, resolved point-source companion of the
central WD, which was argued to be the dM star responsible for the NIR excess.
Remarkably, the location of the $I$-band stellar source coincides with that of
the CEK.

Thus EGB~6 raises several astrophysical puzzles, including how to explain the
existence and survival of a compact dense [\OIII]-emitting nebula apparently
associated with a cool M dwarf, located at least 118~AU from the source of
ionizing radiation. EGB~6 may not be entirely unique, however: Frew \& Parker
(2010) and Miszalski et al.\ (2011) discuss several additional examples of
otherwise normal PNe with compact, unresolved, high-density nebulae at their
centers, comprising a class of ``EGB~6-like'' PNNi. The peculiar emission-line
object Tol~26 (CTIO 1230$-$275), having a high-density, compact nebula (Hawley
1981), may be related to this class.

In this paper, we discuss new \HST\/ images of EGB~6, and present previously
unpublished ground-based photometric and spectroscopic monitoring of the object,
obtained in a search for time variability in the stellar and nebular fluxes. We
also present EGB~6's NIR spectrum, and discuss the spectral-energy distribution
(SED) of the nucleus. We end with some scenarios to explain the origin of this
puzzling object.

\section{An {\em HST\/} Study in Ultraviolet, Optical, Scarlet, and
Near-Infrared}

\subsection{New {\em HST\/} Images}

As noted in \S1, the \HST\/ data discussed by L13 were taken between 1991 and
1995. Two decades later, in late 2013, we obtained new \HST\/ images, for the
purposes of confirming the conclusions of L13, verifying that the CEK is
physically associated with the PNN by showing that they have the same proper
motions, and obtaining NIR images of the source.

Table~1 presents details of our new \HST\/ observations. All data were obtained
with the Wide Field Planetary Camera~3 (WFC3), using its UVIS channel in the
near-ultraviolet (NUV) and optical, and its IR channel in the NIR\null. For the
UVIS imaging, we used a three-point dither pattern in each of seven different
filters: five broad-band filters, extending from the NUV to the \HST\/
equivalent of the $I$ band, and two narrow-band filters covering the emission
lines of [\OIII]~5007~\AA\ and H$\alpha$. The UVIS frames were taken with a
$512\times512$-pixel subarray, yielding a field of view of
$20\farcs3\times20\farcs3$. For the IR imaging, we used a four-point dither
pattern and the F160W filter, with a bandpass similar to the ground-based $H$
band. The IR subarray was also $512\times512$ pixels, giving an angular size of
$65\farcs7\times65\farcs7$. For our analysis, we downloaded the default pipeline
drizzle-combined images in each filter from the Mikulski Archive for Space
Telescopes (MAST)\footnote{\tt http://archive.stsci.edu}. These frames have
cosmic rays removed, are corrected for geometric distortions, and have pixel
values corresponding to counts (electrons) per second.

Figure~1 shows pictorial representations of the WFC3 images.  In the NUV (F225W
and F275W) and optical $U$ (F336W) bands, only the hot DAOZ central star is
seen. The $V$-band image (F555W) shows the companion CEK faintly, but only
because several strong emission lines (principally H$\beta$ and
[\OIII]~4959--5007~\AA) lie within the broad filter bandpass. The narrow-band
images in [\OIII]~5007~\AA\ (F502N) and H$\alpha$ (F656N) clearly resolve the
stellar-appearing CEK from the central star. The widths of the filter bandpasses
for F502N and F656N, according to the WFC3 {\it Instrument Handbook\/} (Dressel
2015), are 65 and 18~\AA, respectively. The much wider bandpass of the [\OIII]
filter compared to that of H$\alpha$ explains why the CEK appears fainter than
the central WD in F502N, but brighter in F656N (in addition to the DAOZ WD
having H$\alpha$ in absorption).  

At the $I$ band (F814W), a faint source is seen at the position of the CEK\null.
We measure this source to be about 3.6~mag fainter than the WD\null.
Unfortunately, we have no spectroscopy of EGB~6 that covers the entire $I$ band.
However, the detected source is almost certainly too bright to be due to
emission lines from the CEK\null. There are very few nebular emission lines in
the F814W bandpass; see, for example, the $I$-band spectrum of the PN IC~2165 in
Fig.~4 of Dufour et al.\ (2015). IC~2165 has a central star with parameters
similar to that of the hot component of EGB~6, according to Henry et al.\
(2015), so a similar nebular emission spectrum is expected. Moreover, DL89
calculated a photoionization model for the EGB~6 CEK, leading to predictions of
emission-line strengths, including lines in the F814W bandpass (their Table~1).
Using either the line fluxes measured in IC~2165 (scaled to the H$\alpha$ flux
in EGB~6), or the model predictions from DL89, as input to the WFC3 Exposure
Time Calculator\footnote{\tt http://etc.stsci.edu/etc/input/wfc3uvis/imaging},
we find that the emission lines can account for no more than $\sim$10\% of the
signal seen in F814W\null. Thus we confirm the argument in L13 that a cool
source of faint $I$-band light lies at the location of the CEK\null. 

At the $H$ band (IR-channel F160W) we see a single source with a stellar
profile. Although the WD is comparable in flux to the cool source in the $H$
band (see \S4 below), the IR channel has considerably larger pixels
($0\farcs128$) compared to those of the UVIS channel ($0\farcs0396$), so we
expect at most a small elongation of the image. However, the two components are
well resolved in ground-based NIR active-optics (AO) data, as discussed below in
\S3.4.


\subsection{Absolute Fluxes}

We used routines from the IRAF\footnote{IRAF is distributed by the National
Optical Astronomy Observatory, which is operated by the Association of
Universities for Research in Astronomy (AURA) under a cooperative agreement with
the National Science Foundation.} software package to carry out aperture
photometry on the WFC3 images. The count rates within apertures with a radius of
$0\farcs4$ (thus including the central star, cool companion, and CEK) were first
determined with the {\tt phot} task. These were then converted to absolute
fluxes, using the photometric zero-points given by the {\tt PHOTFLAM} keywords
in the image headers, and scaling these values (which are for apertures of
infinite radius) to $0\farcs4$ apertures, using ratios derived from the
information given at the WFC3 website\footnote{{\tt
http://www.stsci.edu/hst/wfc3/phot\_zp\_lbn}, accessed on 2015 November~15.} at
Space Telescope Science Institute (STScI).

The resulting $F_\lambda$ values are given in the right-hand columns in Table~1
for the broad-band filters. For the narrow-band filters, we calculated the
emission-line fluxes from the CEK for [\OIII]~5007~\AA\ and H$\alpha$ by
multiplying $F_\lambda$ by the filter bandwidths given above, and then scaling
by the magnitude differences between the central star and the CEK measured in
the two images. The resulting monochromatic line fluxes are given in the final
column of Table~1, for the [\OIII]~5007~\AA\ and H$\alpha$ lines. The line
fluxes are consistent with all of the emission being from the knot.

\subsection{Astrometry}

The separation and position angle (PA) of the CEK with respect to the PNN were
measured by L13 in the \HST\/ frames taken between 1991 and 1995. For
convenience, these measures are reproduced in the first three lines of Table~2.
In order to determine whether our new observations reveal any appreciable
relative motion, we measured the positions of the CEK and PNN, using the
centroiding task in the {\tt imexamine} package in IRAF\null. We then employed
the STSDAS\footnote{STSDAS (Space Telescope Science Data Analysis System) is a
product of STScI, which is operated by AURA for NASA.} {\tt xy2rd} routine to
convert the $x,y$ positions to J2000 right ascension and declination, from which
separation and PA can be determined. The average values from the 2013 frames in
[\OIII] and H$\alpha$ are given in the last line of Table~2. These results show
that there has been no significant change in the separation and PA over the
approximately two-decade interval covered by our observations.

Independent measurements of the absolute proper motion of the central star are
available from the PPMXL (Roeser et al.\ 2010) and UCAC4 (Zacharias et al.\
2013) catalogs, and from Data Release~7 of the Sloan Digital Sky Survey (SDSS)
as quoted by Girven et al.\ (2011). These sources give proper motions,
$(\mu_\alpha, \mu_\delta)$ in $\rm mas\,yr^{-1}$, of $(-11.2, 0.0)$, $(-14.8,
+2.3)$, and $(-12.2, +4.9)$, respectively. Taking the mean of these values, we
find that the total absolute motion of the central star from the epoch of the
first \HST\/ observation to that of the most recent one is $0\farcs29$. Since
the measured separation of the PNN and CEK has changed by no more than a few
mas, this is a strong indication that they are physically associated. There has,
however, been no detectable orbital motion. This is slightly surprising, since
the nominal period, for a true separation of $\sim$118~AU, would be of order
1450~years, and thus about 1/60th of the period would have elapsed between the
two sets of \HST\/ images. This would give a change of $\sim\!6^\circ$ in PA for
a face-on circular orbit, which we do not see. However, there is a wide range of
orbital parameters for a gravitationally bound system that are consistent with
our observations.

\section{Ground-based Data! Data! Data!}

We turn now to ground-based photometry and spectroscopy of the EGB~6 central
object. We discuss absolute photometry, monitoring programs aimed at searching
for photometric and/or spectroscopic variations that might shed light on the
nature of the central source, and an archival but previously unpublished NIR
spectroscopic observation.

\subsection{Absolute Optical Photometry}

We obtained absolute photometry of the central source in EGB~6 on three
photometric nights between 1991 and 1998, using 0.9-m telescopes at Kitt Peak
National Observatory (KPNO) and Cerro Tololo Inter-American Observatory (CTIO),
and calibrated to the Johnson-Kron-Cousins {\it BVRI\/} system via observations
of standard stars from Landolt (1992, 2009). These previously unpublished data
are presented in Table~3, along with results quoted from the literature sources
indicated in the last column. In addition, the last two data entries from 2016
are from observations with the Asiago 0.67/0.92-m Schmidt telescope, and
likewise calibrated to Landolt standards. The first two lines in Table~3 are
from photoelectric observations in 1978 and 1982, with errors of approximately
$\pm$0.05~mag. The remaining data were obtained with CCDs, were reduced with
standard aperture-photometry routines, and have typical errors usually of about
$\pm$0.005--0.010~mag. Averages of the CCD results, and the errors of the means,
are given at the bottom of Table~3. 

The $B$ and $V$ magnitudes of the central object have been essentially constant
within the errors from 1991 to 2016, and, with slightly larger uncertainty,
since the photoelectric observations in 1978--1982. The $R$ magnitude is also
nearly constant, but with perhaps a slightly larger scatter. There does appear
to be variability at the $I$ band, at the level of several hundredths of a
magnitude, even during a single observing run with identical equipment. Based
primarily on the two observations in 1991 and 1994, compared with subsequent
measures, there is a slight suggestion of a slow secular fading of the $I$
magnitude, but this could simply be due to the randomness of the short-timescale
variability.

Our measured magnitude difference of 3.6 between the companion and the WD in the
\HST\/ F814W image (discussed in \S2.1), along with an $I$ magnitude of the
total flux of 16.31, imply that the apparent magnitude of the companion is
$I\simeq19.9$.

\subsection{Differential Optical Photometric Monitoring}

We obtained differential CCD photometry during six different observing runs on
0.9-m telescopes at KPNO (1990 December, 1991 November, 1998 March) and CTIO
(1991 January, 1992 May, 1994 March). These results have likewise not been
published previously. In nearly all cases only a single set of {\it BVRI\/}
observations was made per night during these runs, as they were part of a larger
monitoring survey of many PNNi in search of variable central stars (e.g., Bond
et al.\ 1992; Ciardullo \& Bond 1996). The total number of observations of EGB~6
was 57 in $B$ and $V$, 50 in $R$, and 53 in $I$\null. The seeing in these frames
generally ranged between $1\farcs3$ and $2\farcs2$.

The frames were bias-subtracted and flat-fielded using standard IRAF tasks.
Differential photometry was carried out using the {\tt daophot}
point-spread-function (PSF)-fitting routines within IRAF\null. Each frame's PSF
was determined from three bright, isolated field stars located near the EGB~6
central star. We then calculated the magnitude difference between EGB~6 and the
sum of intensities of these three comparison stars. The formal precisions for
differential magnitudes determined in this way for a source as faint as EGB~6
are about $\pm$0.01~mag in $B$, $V$, and $R$, and about $\pm$0.02~mag in $I$.

%
%
%

Unfortunately, for the purpose of long-term monitoring, this is a cumbersome
batch of data. Not only were the six runs made with five different combinations
of CCD and filter sets, but the PNN ($B-V=-0.31$) is considerably bluer than the
comparison stars (whose combined light has $B-V=0.80$). Thus our differential 
magnitudes are sensitive to changes in the bandpass effective wavelengths among
the different instrumental setups. During three of the six observing runs, 
observations of Landolt standard stars allowed us to determine the color terms,
and thus correct our data to the standard system. For the other three runs (KPNO
1990 December, KPNO 1991 November, and CTIO 1992 May), we simply forced the mean
differential magnitudes to equal the means from the three calibrated runs. (This
would mean that we might miss any systematic long-term changes in brightness,
but the absolute photometry discussed above indicates this is not the case,
except possibly in $I$.)

%
%
%
%
%

Figure~2 plots the resulting magnitudes against date of observation. The
zero-points in each filter have been set to reproduce the mean calibrated
magnitudes given in Table~3, but with $R$ and $I$ offset for clarity by the
amounts indicated in the figure.  The photometric scatter in $B$, $V$, and $R$
is roughly twice that suggested by the photon-statistical errors quoted above,
but this is not surprising because some of the data were taken in
non-photometric conditions, and because of small systematic errors due to
flat-field illumination and other instrumental effects. The scatter in $I$ does,
however, appear to exceed the expected amount; since the absolute photometry
discussed in \S3.1 also suggested variability, we believe that the source is
indeed a short-term variable at the $I$ bandpass. We searched for a periodic
signal using a periodogram analysis, but did not find any evidence for one.
However, the observing cadence of one observation per night was not particularly
suitable for this purpose. An intensive campaign of monitoring would be useful
for further investigation.

Photometry of EGB~6 in the $V$ band is also available from late 2005 to late
2013 from the Catalina Real-time Transient Survey (CRTS)\footnote{{\tt
http://crts.caltech.edu}, accessed 2015 December 23} (Drake et al.\ 2009). No
significant variability at $V$, apart from a few outliers, is seen in more than
700 observations. Typical error bars for the CRTS data points are about
$\pm$0.07~mag.  There are also about 20 photographic observations from the
Digital Access to a Sky Century @ Harvard (DASCH) project (Grindlay et al.\
2012), going back to 1914, again showing no convincing evidence for long-term
variability at blue wavelengths. The star did nothing in the night-time.



\subsection{Optical Spectroscopic Monitoring}

As discussed in \S1, it had been believed that there is an M dwarf companion to
the central WD star of EGB~6. Moreover, the system is suspected of variability
in the $I$ band (and in the NIR---see below). This suggested the possibility
that the cool component could be a dM flare star or other type of variable. If
so, the optical spectrum during a flare might exhibit enhanced Balmer emission
as well as emission at \ion{Ca}{2} H and K\null. Moreover, the discussion in L13
raised the possibility that the electron density in the CEK could be changing,
for example if the dense compact nebula is dispersing. In this case, the nebular
forbidden emission-line spectrum might be variable.

With these motivations, we monitored the central source by arranging to obtain
optical spectra with the 1.5-m telescope at CTIO operated by the SMARTS
Consortium\footnote{SMARTS is the Small \& Moderate Aperture Research Telescope
System; {\tt http://www.astro.yale.edu/smarts}}. The observations were 
conducted by Chilean service observers on 47 nights between 2004 January~31 and
2012 January~24. We used the RC-focus spectrograph equipped with a CCD camera,
and grating 26 in first order, covering either 3532--5300~\AA\ or
3660--5440~\AA, at a spectral resolution of 4.3~\AA\null. Exposure times each
night were $3\times400$~s. A short exposure on a HeAr lamp was taken before each
set of stellar observations for wavelength calibration. The CCD images were
bias-subtracted and flat-fielded, combined for cosmic-ray removal, and then the
spectrum was extracted and wavelength-calibrated, all using standard IRAF
routines. The spectra were normalized to a flat continuum, and smoothed with a
three-point boxcar kernel.

The top panel in Figure~3 shows two examples of the spectra, obtained near the
beginning and end of the eight-year observing interval, in order to illustrate
the quality of the individual data. This panel also shows a mean spectrum made
by combining all 47 observations. The strongest emission lines are marked. The
bottom panel in Figure~3 shows the same combined spectrum, with the vertical
scale expanded so as to show the photospheric absorption features from the WD
more clearly. 

We see no strong evidence for spectroscopic variability in these data. Enhanced
Balmer emissions, and the appearance of \ion{Ca}{2} H and K in emission, never
occurred. Using standard IRAF tasks, we combined the spectra into seasonal
averages, and determined the equivalent widths (EWs) of the emission lines
during each season. We also measured the EWs in the combined spectrum of all 47
observations. The results are presented in Table~4. The uncertainties for the
seasonal averages are of order a few percent for the strong lines, and up to
about 12\% for the weaker ones. In the next-to-last line of Table~4 we list the
EWs published by L89, based on the averages of seven spectra obtained with
various telescopes over the interval 1978 to 1987. We see no evidence for large
changes in the emission-line spectrum over the entire range from 1978 to 2012.
(There is a slight suggestion that [\OIII] 5007~\AA\ and H$\beta$ have weakened,
but at least for the latter this could be a systematic effect due to the blend
with the photospheric absorption feature, combined with the relatively low
spectral resolution of the SMARTS spectra.)

As this paper was being completed, we were able to obtain spectrograms of EGB~6
on 2016 April~3, using the Asiago Astrophysical Observatory 1.22-m telescope.
The exposures were $3\times1200$~s. Equivalent widths derived from these spectra
are given in the final entry in Table~4. They are generally consistent with the
values in the rest of the table, showing again that any secular changes in the
emission-line spectrum since the late 1970s have been small. The new observation
provides marginal evidence for a slow strengthening of [\OIII] 4363~\AA, which
if so would suggest an increase in the electron density. There is further
evidence that H$\beta$ has weakened, and [\OIII] 4959--5007~\AA\ have
strengthened. Continued monitoring of the spectrum is desirable.

%
%
%

\subsection{Gemini Near-IR Spectrum}

In a search of the Gemini Observatory Archive\footnote{\tt
https://archive.gemini.edu}, we found that observations of the central source in
EGB~6 had been obtained with the 8-m Gemini North NIR spectrograph (GNIRS; Elias
et al.\ 2006) on 2012 December~29 (program GN-2012B-Q-60; PI:
J.~Bil{\'{\i}}kov{\'a}). The observing sequence was a standard ABBA pattern of
$8\times225$~s exposures on EGB~6 and $6\times4$~s exposures on HR~4041, a
neighboring A0~V star to be used for telluric correction. GNIRS was operated
using AO in its cross-dispersed mode with the $10\,\rm lines\, mm^{-1}$ grating
and a $0\farcs10$ slit. Wavelength coverage was 0.95--2.5 $\mu$m, at resolution
$R = 5000$.  

In the $H$-band acquisition images, we noticed that EGB~6 was partially resolved
into two point sources, at a separation consistent with the \HST\/ observations
described above. The spectrograph slit, perhaps fortuitously, was oriented such
that it lay almost exactly along the nearly east-west orientation of the two
sources; thus the spectra of the two sources are spatially separated. Visual
examination shows that the hot WD dominates the continuum emission blueward of
$\sim$$1.25\,\mu$m. At longer wavelengths, the cool source brightens and the WD
weakens. In the $H$ band, the two sources have nearly equal fluxes. The cool
companion dominates redward of there. 

Figure~4 presents close-ups of four sections of the two-dimensional spectrum.
The first panel on the left shows the region around the \HeI\ 10830~\AA\
emission line from the CEK\null. The emission is spatially offset from the
continuum of the WD, consistent with our findings from the narrow-band \HST\/
images discussed in \S2.1 and in L13. The second panel is centered on the region
around Paschen-$\beta$ at 12818~\AA; this emission line is likewise offset from
the WD continuum, and coincides instead with the weakly visible continuum of the
cool source. The third panel lies near the center of the $H$ band; now the
continua of the WD and cool source have similar brightnesses. In the final
panel, near the center of the $K$ band, the spectrum of the cool source is
brighter than that of the WD. 

After removing electronic artifacts from the images using the {\tt cleanir.py}
routine provided by the Gemini Observatory, we processed the data using standard
IRAF tasks for long-slit spectroscopy. Consecutive images were first
flat-fielded, then subtracted to remove sky emission, and finally stacked. Due
to the semi-resolved nature of the sources, we elected to extract the combined
spectrum of both objects using an aperture with a width of
12~pixels\footnote{This extraction width corresponds to an angular width of
$0\farcs62$. We attempted to extract separate spectra of the two sources, but
were only partially successful, due to the varying flux ratio between the
sources and across spectral orders.}. The spectrum for HR~4041 was processed in
a similar manner. An argon arc-lamp spectrum taken immediately after the science
observations provided wavelength calibration.  To remove telluric absorption
lines and correct for instrumental response, we divided the combined spectrum of
the two EGB~6 sources by the spectrum of HR~4041, after first interpolating
across its intrinsic hydrogen absorption lines. This ratio spectrum was then
multiplied by the $F_\lambda$ vs.\ wavelength relation for a 9500~K blackbody
(the approximate effective temperature of HR~4041). The zero-point flux level of
the final spectrum was normalized to match the available absolute NIR
photometry (tabulated below in Table~5). 

Figure~5 plots the resulting spectrum of the WD plus companion source as a black
line. We saw no absorption features in the spectrum at its modest SNR, so we
applied an 11-point boxcar smoothing. Several prominent emission lines from the
CEK are labeled\footnote{A strong emission line is detected at $2.436\,\mu$m,
which we have been unable to identify. Inspection of the spectrum image
indicates that it appears to be real, and is spatially associated with the
companion source.}. To recover the contribution of the cool companion, we
subtracted a 93,230~K blackbody, representing the WD component, shown as a blue
curve. The resulting companion spectrum is plotted as a red line.

We stared at it in astonishment: the companion spectrum is {\it not\/} that of
the M dwarf that has been claimed by earlier authors (\S1), including our own
previous report in L13. Instead, longward of about $1.3\,\mu$m, there is a
smooth continuum, consistent with that of a cool blackbody. There appears to be
a broad bump from approximately 1.0 to $1.2\,\mu$m, but this feature may be of
doubtful reality since the intrinsic signal in this part of the spectrum (before
conversion to $F_\lambda$) is quite weak and noisy. (If the bump is real, we
have no obvious explanation for it; for example, it is much too narrow to be
attributed to a blackbody contribution, and there are no features in late-type
stellar spectra with such a structure.) Overall, the companion's energy
distribution can be fitted approximately by a blackbody of about 1850~K, plotted
as a green line in Figure~5, or somewhat cooler if we were to discount the
1.0--$1.2\,\mu$m bump. 

Thus the Gemini spectrum has clearly revealed the spatial location of the NIR
excess that was discovered more than two decades ago. The NIR source is a
point-like companion of the central star, which coincides with the unresolved
CEK seen in the \HST\/ images. It is not a dM star, but a considerably cooler
source.

\section{The Sign of the Four: The Spectral-Energy Distribution}


\subsection{Broad-band Photometry and Calibrated Spectra}

In Table~1 we had presented the continuum fluxes ($F_\lambda$) derived from our
2013 \HST\/ broad-band photometry. In Table~5 we collect additional broad-band
photometry of the EGB~6 central source and accompanying CEK from the following
archival and literature sources: (1)~FUV and NUV magnitudes from the {\it Galaxy
Evolution Explorer\/} (\GALEX); (2)~a $U$ magnitude from D15; (3)~mean {\it
BVRI\/} magnitudes from our Table~3 (which included data from D15); (4)~{\it
ugriz\/} photometry from the SDSS; (5)~the mean of the {\it JHK\/} magnitudes
listed by FL93; (6)~the 2MASS $JHK_s$ magnitudes; (7)~the \WISE\/ $W1$ through
$W4$ magnitudes; and (8)~fluxes from 3.6 to $24\,\mu$m from the \Spitzer\/ IRAC
and MIPS instruments. The footnotes to Table~5 give literature or archive
references for each of these sources of data. Column~4 gives effective
wavelengths for each bandpass, with the literature sources given in another
footnote. The magnitudes have then been converted to $F_\lambda$ values in the
fifth column, using zero-points referenced in the final footnote. 

We also make use of {\it Herschel Space Observatory\/} imaging of EGB~6,
available at the {\it Herschel\/} Science Archive\footnote{\tt
http://www.cosmos.esa.int/web/herschel/science-archive}.  The target was
observed (Program OT1\_ksu\_2; PI: K.~Su) with the Photodetector Array Camera
and Spectrometer (PACS; Poglitsch et al.\ 2010) and the Spectral and Photometric
Imaging Receiver (SPIRE; Griffin et al.\ 2010) on 2011 November 27 and 2011
October 23, respectively. The PACS observations consist of images at 70 and
160~$\mu$m, but the central source in EGB~6 was only detected at 70~$\mu$m.
EGB~6 was likewise undetected in the SPIRE data (250, 350, and 500~$\mu$m). We
reduced the PACS data using the {\it Herschel\/} Interactive Processing
Environment (HIPE; Ott 2010) software version 9.0.0, following the standard
pipeline. We measured the flux density at 70~$\mu$m using aperture photometry
with a radius of 6$\farcs$0, and then applying the aperture corrections
suggested by the NASA {\it Herschel\/} Science Center\footnote{\tt
https://nhscsci.ipac.caltech.edu/sc/index.php/Pacs/ApertureCorrections}. Since
the error maps created by the HIPE are unreliable, we estimated the flux error
by taking the standard deviation of the background measurements in five
arbitrarily distributed  6$\farcs$0 apertures placed in the high-coverage area
of the observations. We also estimated an upper limit to the non-detection at
160~$\mu$m in a similar manner and report the measured standard deviation
multiplied by three. The final results are flux densities of $3.6\pm0.6$~mJy at
70~$\mu$m, and an upper limit of 15~mJy at 160~$\mu$m. The 70~$\mu$m value,
converted to $F_\lambda$, is given in the final line of Table~5.

In addition to the broad-band photometry presented in Table~5, and the Gemini
NIR spectrum discussed in \S3.4, there are flux-calibrated space-based
spectroscopic observations available from the following archival sources:
(1)~three FUV spectra (1150--1970~\AA) obtained with the {\it International
Ultraviolet Explorer\/} (\IUE) in 1987 (PI: J.~Holberg) and 1989 (PI: J.L.);
(2)~two spectra, covering 1140--2500~\AA\ and 2220--3300~\AA, obtained with the
Faint Object Spectrograph (FOS) on \HST\/ in 1992 (PI: H.~Shipman); and (3)~a
MIR spectrum (5.25--37.3~$\mu$m) obtained with the Infrared Spectrograph (IRS)
on \Spitzer\/ in 2009 (PI: K.~Su). We downloaded these data from MAST (for
\IUE\/ and FOS) and from the Cornell Atlas of \Spitzer\/ IRS
Sources\footnote{\tt http://cassis.sirtf.com/atlas} (for IRS; Lebouteiller et
al.\ 2011).

\subsection{Variability}

In Figure~6 we plot $\lambda F_\lambda$ values based on Tables~1 and 5, and the
spectroscopic fluxes, against wavelength. (This figure is an update of one
presented in a conference poster by Su et al.\ 2011\footnote{The poster is
available at {\tt http://www.jb.man.ac.uk/apn5/poster\_pdfs.html}}.) The legend
in the figure identifies the sources of the plotted photometry and spectra. From
the FUV through the NUV and optical, up to about the $R$ band, all of the data
are in excellent agreement, in spite of having been obtained over a wide
interval of dates. We thus verify again that EGB~6 is non-variable over this
spectral range.  We had already noted evidence for variability in the
ground-based $I$ band in \S3.2. Figure~6 shows that even larger discrepancies
start to appear as we move to longer wavelengths. The NIR and MIR data were
obtained over a range of dates: in chronological order (1)~the ground-based
$JHK$ observations were made in 1991 (FL93), with variations on a timescale of
one day noted; (2)~the 2MASS data, especially discrepant with FL93 at $K$, are a
combination of survey observations obtained from 1997 to 2001; (3)~the four
\Spitzer\/ IRAC data points---which are systematically bright by $\sim$0.75~mag
relative to the rest of the observations---were obtained in 2007, and the MIPS
datum in 2008; (4)~the \Spitzer\/ IRS spectrum is from 2009; (5)~the four
\WISE\/ points are from survey observations made over seven months in 2010;
(6)~the {\it Herschel\/} 70~$\mu$m datum was obtained in 2011; and (7)~the
Gemini NIR spectrum is from 2012.

\subsection{The Spectral-Energy Distribution}

Because of the variability of the EGB~6 nucleus in the NIR and MIR, and the fact
that the data plotted in Figure~6 were obtained at a range of different epochs,
it is doubtless premature to attempt a general, static model to explain the
SED\null. However, as a rough guide to the astrophysical parameters, here we
will simply fit the SED of the combined light of the nucleus and companion with
four components, represented by blackbodies. We fixed the temperature of the hot
WD at 93,230~K from the atmospheric analysis (\S1). The optical colors as well
as the \IUE\/ and \HST/FOS spectra show that the WD is very lightly reddened; we
adopted $E(B-V)=0.02$, giving a good fit to the SED from the FUV to the optical.
Since the total reddening in the direction of EGB~6 is $E(B-V)\simeq0.027$,
according to the reddening maps of Schlafly \& Finkbeiner (2011)\footnote{as
implemented at the NASA/IPAC Extragalactic Database, {\tt
http://ned.ipac.caltech.edu\slash forms\slash calculator.html}}, it appears that
the hot central star suffers no significant reddening within the system.

We arbitrarily omitted the four high \Spitzer\/ IRAC points\footnote{ To verify
the C11 results that produced the four high values, we obtained the IRAC data
from the \Spitzer\/ archive ({\tt http://irsa.ipac.caltech.edu}), combined the
individual corrected basic calibrated data images into a mosaic, and performed
aperture photometry using the Mosaicking and Point Source Extraction package
developed by the \Spitzer\/ Science Center (Makovoz \& Marleau 2005). Our
results are in agreement with those of C11.}, because the remaining data are
reasonably self-consistent, although still showing some discordances due to
variability, and then performed a $\chi^2$ fit of four blackbodies to the
remaining data. The blackbodies were corrected for the adopted reddening, using
the formulae of Cardelli et al.\ (1989) with $R_V=3.1$. The resulting SEDs of
these blackbodies are shown as dashed black lines in Figure~6, and their sum as
a solid red line. The MIR data are fit reasonably well by two cool blackbodies,
corresponding to dust with temperatures of 385 and 175~K\null. This result is in
fairly good agreement with C11, who found 500 and 150~K for these components,
but using only the \Spitzer\/ IRAC and MIPS data. However, these two
blackbodies, combined with the Rayleigh-Jeans tail of the WD, fail to account
for the excess flux in the NIR, whose presence as a separate source is clearly
revealed by the Gemini spectrum. This NIR excess has been attributed to an
M3--M5 dwarf ($\Teff\simeq3400$--3000~K\footnote{For nominal stellar parameters
as functions of spectral type, here and in the next section, we use a literature
compilation assembled by E.~Mamajek: \tt http://www.pas.rochester.edu\slash
$^\sim$emamajek\slash EEM\_dwarf\_UBVIJHK\_colors\_Teff.txt}) in previous
studies, as summarized in \S1 and in L13 and references therein. However, our
$\chi^2$ fit, as well as the Gemini spectrum (\S3.4 and Figure~5), require a
much lower temperature for this component, around 1850~K\null.  

\section{Dust is an Essential Part of the System}

To summarize the results presented here and in L13: the hot WD nucleus of the
old PN EGB~6 is accompanied by a compact source at a projected separation of
$\sim$118~AU\null. This companion object emits both forbidden emission lines
from a dense nebula, and NIR flux with a blackbody-like spectrum at a
temperature of  $\sim$1850~K\null. The NIR flux varies on a timescale possibly
as short as a few days, but the nebular emission spectrum has not shown any
large variability. In addition, there is cooler dust, with temperatures of
175--385~K, whose physical location within the system remains unknown from the
available observations. The flux from at least the 385~K component of this dust
is also time-variable.

L13 discussed two scenarios that might explain the emission-line knot seen in
EGB~6, based on their conclusion that it is physically associated with an
M~dwarf: (1)~the CEK is a stable region of compressed gas where winds from the
WD and dM star collide, or (2)~the CEK is a photoionized remnant envelope or
accretion disk around the dM star, which (as first suggested by Zuckerman et
al.\ 1991) was captured during the epoch of rapid mass outflow from the WD
progenitor that created the large, faint surrounding PN a few times $10^4$~yr
ago. Our new finding that there is {\it not\/} an exposed M dwarf in the system
now makes the colliding-wind scenario unlikely. 



A combination of four blackbodies---WD, NIR source, and the two cool MIR
components---does fit the entire SED reasonably well (again, apart from the four
discrepantly high \Spitzer\/ IRAC points), as depicted by the red line in
Figure~6.  L13 estimated a distance to EGB~6 of $d=576^{+1224}_{-271}$~pc, based
on the absolute magnitude of the WD component derived from its atmospheric
parameters; the large uncertainty in $d$ is due to relatively large
uncertainties in the $\Teff$ and $\log g$ of the WD, a consequence of
emission-line contamination of the photospheric Balmer lines in the spectra
analyzed by G10. However, Frew et al.\ (2016) have developed a statistical
distance indicator for PNe based on an H$\alpha$ surface-brightness vs.\
physical radius relation. Applying this calibration to the surrounding large,
faint PN of EGB~6, Frew et al.\ find a similar distance but with a smaller
uncertainty: $d=870\pm250$~pc. 

If the NIR source were a star with an effective temperature  of $\sim$1850~K, it
would be an early L-type dwarf, with a $K$-band absolute
magnitude of roughly +11.3. At a distance of about
725~pc (the unweighted average of the above two estimates), this L dwarf would
have an apparent magnitude of $K\simeq20.6$; however, the observed $K$
magnitudes (Table~5) are in the range 15.6--16.1. Moreover, the observed NIR
spectrum does not have the strong molecular bands seen in L-type dwarfs; as
described in \S3.4, it has a smooth blackbody-like spectrum longward of
$\sim$1.2~$\mu$m. Clearly the NIR source is not a stellar photosphere.



%
%
%
%
%
%



At a separation of at least $\sim$118~AU from the WD, the NIR source is much too
warm to be heated by the WD\null. The NIR source is more plausibly a dusty
envelope or disk, which reradiates the luminosity of a heavily obscured low-mass
star. The physical radius of the 1850~K NIR source, for a 725~pc distance, is
about $0.86\,R_\odot$. Its radiated luminosity is $\log L/L_\odot\simeq-2.1$. If
this luminosity is due entirely to reprocessed radiation from an enshrouded
star, it corresponds to that of an M3.5~V dwarf, using the Mamajek table cited
above. In this picture, the companion source is an M dwarf after all, but one
that is hidden from view by surrounding dust\footnote{Alternatively, the
putative enshrouded companion could be a WD with the required luminosity,
although it requires ``fine-tuning'' of the time since its formation; if so, it
would currently have an effective temperature of about 15,000~K and a cooling
age of $\sim\!2\times10^8$~yr. This is based on an assumed mass of
$0.6\,M_\odot$ and the ``Montreal'' WD cooling tracks at {\tt
http://www.astro.umontreal.ca/$^\sim$bergeron/CoolingModels} (e.g., Tremblay et
al.\ 2011).}. 

In the remainder of this section we discuss two alternative pictures: one in
which the NIR and MIR sources have separate locations in the system, and another
in which they are both emitted by the obscured companion.

\subsection{Dusty Debris Cloud?}

We first consider a scenario in which the source of the MIR excess in EGB~6 does
not coincide with the NIR and emission-line companion, but is located elsewhere
in the binary system. Over the past decade, it has been discovered that a
significant fraction of hot WDs and PNNi are associated with NIR and/or MIR
excesses, indicating the presence of warm and/or cool dust. Su et al.\ (2007,
hereafter S07), using \Spitzer\/ observations, detected a MIR flux from the
central star of the Helix Nebula (NGC~7293), a well-known and very nearby
PN\null. This MIR source exhibits a thermal continuum, with a blackbody
temperature of $\sim$120~K, attributed to an optically thin debris disk of cool
dust surrounding the hot WD at separations of 35--150~AU\null. S07 suggest that
this dust arose from collisions of Kuiper-Belt objects or breakup of Oort-Cloud
comets, due to the dynamical perturbation induced by the sudden mass loss from
the central star when the PN was ejected. \HST\/ imaging of the Helix PNN
(Ciardullo et al.\ 1999) detected no companion stars with spectral types earlier
than M8 (for projected separations greater than 65~AU) to M5 (for separations as
close as $\sim$11~AU). S07 noted moreover that the Helix PNN shows no NIR
excess; thus there is no compelling argument that it has a binary companion.


Although the \HST\/ and NIR observations of EGB~6 clearly associate the NIR
excess (and the emission-line source) with a companion object, there is no
direct evidence that the MIR excess is located at the position of the companion.
Thus the Helix results may suggest a scenario in which the MIR excess in EGB~6
is likewise due to a debris disk encircling the hot WD nucleus, and is not
directly associated with a companion star

Searches for IR excesses associated with hot WDs and PNNi have been made by C11
(see \S1), Bil{\'{\i}}kov{\'a} et al.\ (2012), and Clayton et al.\ (2014,
hereafter C14). (Hoard et al.\ 2013 carried out a wider-scale search of the
all-sky \WISE\/ archive for dust excesses around a large sample of WDs over a
broad range of temperatures, but excluded known PNNi and binaries, including
EGB~6.)  C11 found nine cases of MIR excesses out of a sample of 71 hot WDs;
seven of the nine WDs are PN nuclei, including EGB~6 itself.  Dust temperatures
are typically 120--190~K, but four of the PNNi---which C11 call
``EGB~6-like''--- also exhibit NIR excesses implying additional, warmer dust
components with temperatures of 500--1200~K\null. Bil{\'{\i}}kov{\'a} et al.\
searched the \Spitzer\/ archive and found additional examples; overall, about
18\% of PNNi have associated dust disks. This is a significantly higher fraction
than the $\sim$1--3\% incidence of NIR excesses due to warm dust around cool
($<$25,000~K) WDs (Farihi et al.\ 2009; C14 and references therein). A further
difference is that the warm-dust disks around the cool WDs lie very close to the
star, typically within the tidal-disruption radius for asteroidal bodies. As
discussed by C14, this suggests that the dust around PNNi is recently formed and
relatively transitory, in addition to lying at much greater distances from the
central stars as demanded by the low dust temperatures. These authors note that
eight out of 13 PNNi with dust disks detected at 8 or $24\,\mu$m are known or
suspected to have binary companions. This raises the possibility that a binary
interaction with the asymptotic-giant-branch (AGB) wind aids the formation of a
dusty disk around the mass-losing star in at least some systems, if not in the
Helix Nebula.

Stone et al.\ (2015, hereafter SML15) have discussed these phenomena from a
theoretical standpoint. They propose that the cool dust debris at large
separations from PNNi arises from the response of a pre-existing Oort-Cloud
analog to the sudden ejection of the PN and the ``natal kick'' received by the
WD\null. Some of the comets would be placed in orbits bringing them close enough
to the WD for evaporation or even tidal disruption, leading for formation of a
spherical, optically thin, cool dust cloud around the WD\null. Their model does
not directly account for the warmer dust detected in the NIR---but, at least in
the case of EGB~6, this dust appears to be associated with a companion rather
than a cloud around the PN nucleus. For EGB~6 itself, modeling by SML15 finds
that the $24\,\mu$m flux implies a dust mass of $\sim\!365\,M_\earth$.

\subsection{Post-Mira Accretion Disk?}

Now we consider an alternative scenario in which the EGB~6 nucleus is the
immediate descendant of a symbiotic-like binary similar to the Mira (o~Ceti)
system. Mira itself, a prototypical mass-losing long-period variable star, is
accompanied by a companion, Mira~B, at a separation (in 1995) of $0\farcs578$
(Karovska et al.\ 1997), corresponding to a projected linear separation of
$\sim$50~AU\null. Mira~B is hidden by an optically thick accretion disk of
material captured from the wind of Mira~A (e.g., Ireland et al.\ 2007, hereafter
I07, and references therein). I07 argue on both theoretical and observational
grounds that the underlying Mira~B star is a late-type dwarf, although many
other authors (e.g., Sokoloski \& Bildsten 2010 and references therein) cite
evidence for it being a WD.

Based on the statistics of wide binaries, I07 predict that binary systems in
which an accretion disk is formed around a companion during the AGB wind phase
are relatively common---about one in five among solar-type stars. Once the AGB
star becomes a WD and its slow, dusty stellar wind dies out, the companion's
accretion disk will also eventually disappear. Citing Alexander et al.\ (2006)
for the timescale for viscous evolution of such accretion disks, I07 conclude
that ``systems like Mira should produce clear observational signatures of an
accretion disk around the secondary for at least a few times $10^5$~yr after the
primary becomes a WD.''

Wind accretion from AGB stars in wide binaries has been studied theoretically by
a number of authors, including Soker \& Rappaport (2000), Perets \& Kenyon
(2013, hereafter PK13), and Huarte-Espinosa et al.\ (2013). PK13 state that
``for separations of 3--100~AU $\dots$ wind-fed disks have surface-density and
temperature profiles similar to those observed in low-mass protoplanetary
disks.'' PK13 even raise the possibility of planet formation within such
wind-created accretion disks. 


If we attribute both the NIR and MIR flux to the companion object in EGB~6, then
in fact it does have several features in common with young stellar objects
(YSOs), in particular those of class 0/I (the youngest protostars, with both an
obscuring envelope and an accretion disk). These traits include: (1)~significant
NIR and MIR excesses; (2)~large-amplitude variability in the NIR and MIR (up to
2~mag in extreme cases, e.g., Carpenter et al.\ 2001; Morales-Calder\'on et al.\
2011); and (3)~occasionally H, He, and other species in emission, including
[\OIII] (e.g., van Loon et al.\ 2010). These phenomena arise from combinations
of accretion, disk morphology, shocks, starspots, and ionizing radiation from
nearby bright stars. These give rise to a wide variety of NIR and MIR SEDs and
spectra of YSOs (e.g., Furlan et al.\ 2011). 

For a direct comparison of the NIR/MIR SED of the EGB~6 companion with those of
YSOs, in Figure~7 we plot its position in color-magnitude and color-color
diagrams for combinations of the 2MASS $J$ and \WISE\/ magnitudes. The
contribution of the hot WD to the EGB~6 magnitudes has been removed. We also
plot data for samples of YSOs in the Taurus and Upper Scorpius star-forming
regions, taken from Esplin et al.\ (2014, hereafter E14) and Luhman \& Mamajek
(2012), respectively. The $W1$ magnitudes for EGB~6 and Upper Scorpius have been
adjusted to the 140~pc distance of the Taurus region.

Figure~7 illustrates that the infrared luminosity and colors of the EGB~6
companion are consistent with those observed in YSOs. In this interpretation,
the NIR/MIR excess, SED, and spectrum of EGB 6 are attributed to a companion
object, embedded in an accretion disk and envelope. The star itself is obscured,
and the disk and envelope, which have a wide range of effective temperatures,
provide the IR excess.  The emission lines of H and possibly He would be
produced by accretion shocks from in-falling material (e.g., Calvet \& Hartmann
1992), while the forbidden lines are formed in an outflow or wind. Variations in
the accretion rate could be responsible for the observed infrared variability of
the system.


In Figure~7, we highlight (filled green circles) a particular YSO, Haro~6-39 in
the Taurus region. We chose this source because, as the figure shows, it has
very similar colors to EGB~6 (but is somewhat more luminous).  E14 report that
NIR spectra of Haro~6-39 show no absorption features, and it has H and He
emission (again, a signature of accretion). Thus this YSO has several properties
remarkably similar to those of the EGB~6 companion.

The direction of evolution of YSOs in Figure~7 is from right to left, i.e., from
young, very red class 0/I protostars on the right, through class II and then to
the least obscured and bluest class III objects on the left. The EGB~6 companion
has the characteristics of a relatively young object---consistent with our
interpretation of a recent accretion process---but with a very different origin
from the true YSOs in Taurus and Scorpius. EGB~6 does have a luminosity, as
represented by the $W1$ magnitude, that is relatively low compared to most YSOs
at the same $W1-W2$ color, which may reflect its different formation process.

There is an extensive literature on much closer binary systems among post-AGB
stars. There is substantial evidence for interactions with the companion stars
during the AGB mass-losing phase. Recently, for example, a dusty circumstellar
disk has been directly imaged around the post-AGB binary IRAS~08544$-$4431 by
Hillen et al.\ (2016). They have moreover detected a likely compact accretion
disk around the binary companion in this system. However, with an orbital period
of only 499~days, the companion is at least two orders of magnitude closer to
the mass-losing star than in the case of EGB~6.

\section{The Final Problems}

To summarize, we have considered two scenarios to explain the astrophysical
puzzles presented by EGB~6.
 
(1)~One interpretation is that the MIR excess is due to a large-scale, optically
thin dust disk surrounding the WD, making it an analog of the cool dust disks
seen around a significant fraction of PNNi. In this case, the faint companion
star is surrounded only by warm dust, located very close to the companion, and
having an effective temperature of $\sim$1850~K, which accounts for the NIR
excess flux. Evaporation from the outer surface, and photoionization due to
illumination by the nearby hot WD, produce the compact emission-line nebula. To
account for the optically thick dust shell around the companion star, we could
speculate that it was produced by pulverized rocky bodies that passed very close
to the M~dwarf. In fact, the inferred radius of this shell,
$\sim\!0.86\,R_\odot$, is quite close to the tidal-disruption radius of a
low-mass star.

(2)~Alternatively, the central hot WD has a companion M dwarf (or possibly a
second WD), which captured an optically thick accretion disk during the
mass-ejection episode that produced the surrounding large PN a few times
$10^4$~yr ago. In this picture, the system is descended from a binary similar to
the present-day Mira---that is, a mass-losing AGB star accompanied by a distant
main-sequence (or WD) companion. Both the NIR and MIR excesses of EGB~6 are due
to this thick accretion disk. Its properties, including short-timescale
variability, are similar to those of YSOs in star-forming regions. A portion of
the accretion disk is evaporating rather than falling onto the companion, or
there are possibly jets being ejected. Photoionization of the ejecta by the UV
flux of the PNN produces the forbidden-emission-line source. 

A key uncertainty in distinguishing between these possibilities is the currently
unknown physical location of the MIR source. High-resolution imaging, or even
just high-precision astrometry, with the James Webb Space Telescope might be
able to make this distinction. Another useful program would be simultaneous NIR
and MIR photometric monitoring of the EGB~6 nucleus. If both the NIR and MIR
fluxes vary in tandem, it would argue strongly that they are located close
together; but if the NIR variations are not associated with variations in the
MIR, it could indicate physically separate locations.

Also useful would be to apply investigations as comprehensive as the ones
described here to a larger sample of PNNi with IR excesses. Such studies would
help us understand empirically whether all of the EGB~6-like central stars have
binary companions, or whether these phenomena can arise even from single PN
nuclei. As the fictional Sherlock Holmes pointed out, ``it is a capital
mistake to theorize before you have all the evidence.''

\acknowledgments

Partial support for this work was provided by NASA through grant number
GO-13469 from the Space Telescope Science Institute, which is operated
by AURA, Inc., under NASA contract NAS 5-26555. 

We thank the STScI Director's Discretionary Research Fund for supporting our
participation in the SMARTS consortium, and Fred Walter for scheduling the
1.5-m  queue observations. We especially appreciate the excellent work of the
CTIO/SMARTS service observers who obtained the direct images and spectra during
many long clear Tololo nights:
Claudio Aguilera,
S.~Gonz\'alez, 
Manuel Hern\'andez, 
Rodrigo Hern\'andez,
Alberto Miranda,
Alberto Pasten,
and Jos\'e Vel\'asquez.

The DASCH project at Harvard is grateful for partial support from NSF grants
AST-0407380, AST-0909073, and AST-1313370.

This research has made use of the SIMBAD database, operated at
CDS, Strasbourg, France.

This research has made use of the NASA/IPAC Infrared Science Archive, which is
operated by the Jet Propulsion Laboratory, California Institute of Technology,
under contract with NASA. 

This publication makes use of data products from the Two Micron All Sky Survey,
which is a joint project of the University of Massachusetts and the Infrared
Processing and Analysis Center/California Institute of Technology, funded by
NASA and the NSF. 

It also makes use of data products from the Wide-field Infrared Survey
Explorer, which is a joint project of the University of California, Los Angeles,
and the Jet Propulsion Laboratory/California Institute of Technology, and
NEOWISE, which is a project of the Jet Propulsion Laboratory/California
Institute of Technology. WISE and NEOWISE are funded by NASA.

This work is based in part on observations made with the {\it Spitzer Space
Telescope}, which is operated by the Jet Propulsion Laboratory, California
Institute of Technology under a contract with NASA.

The Cornell Atlas of \Spitzer/IRS Sources (CASSIS) is a product of the 
Infrared Science Center at Cornell University, supported by NASA and JPL. 

Based in part on observations obtained at the Gemini Observatory, which is
operated by the Association of Universities for Research in Astronomy, Inc.,
under a cooperative agreement with the NSF on behalf of the Gemini partnership:
the National Science Foundation (United States), the National Research Council
(Canada), CONICYT (Chile), Ministerio de Ciencia, Tecnolog\'{i}a e
Innovaci\'{o}n Productiva (Argentina), and Minist\'{e}rio da Ci\^{e}ncia,
Tecnologia e Inova\c{c}\~{a}o (Brazil).  

{\it Herschel\/} is an ESA space observatory with science instruments provided
by European-led Principal Investigator consortia and with important
participation from NASA. 

We thank S.~Dallaporta for his assistance with the Asiago observations.


{\it Facilities:} 
\facility{IUE},
\facility{{\it Hubble Space Telescope} (STIS, WF/PC1,
WFPC2, WFC3)},
\facility{CTIO:0.9-m, CTIO:1.5m},
\facility{Gemini},
\facility{Spitzer},
\facility{WISE},
\facility{Herschel}

\newpage

\begin{deluxetable}{cclcc}
\tablecaption{New {\em HST\/} Observations of EGB~6} 
\tablewidth{0pt}
\tablehead{
\colhead{Date} &
\colhead{Camera} &
\colhead{Filter} &
\colhead{Total } &
\colhead{$F_\lambda$ or $I_\lambda$\tablenotemark{a}} \\
\colhead{} &
\colhead{} &
\colhead{} &
\colhead{Exposure [s]} &
\colhead{}
}
\startdata 
2013 Dec 6 & WFC3/UVIS & F225W &  15    & $3.20\times10^{-14}$ \\
$''$       & $''$      & F275W &  15    & $2.02\times10^{-14}$ \\
$''$       & $''$      & F336W &  15    & $9.05\times10^{-15}$ \\
$''$       & $''$      & F502N &  360   & $4.2\times10^{-14}$ \\
$''$       & $''$      & F555W &  15    & $1.66\times10^{-15}$ \\
$''$       & $''$      & F656N &  720   & $2.0\times10^{-14}$ \\
$''$       & $''$      & F814W &  30    & $3.44\times10^{-16}$ \\
$''$       & WFC3/IR   & F160W &  20.47 & $4.23\times10^{-17}$ \\
\enddata
\tablecomments{Observations were made in program GO-13469 (PI: H.E.B.)}
\tablenotetext{a}{For the broad-band filters, this column contains the total
flux  density ($F_\lambda$ in $\rm erg\,cm^{-2}\,s^{-1}\,\AA^{-1}$) measured for
the central source. For [\OIII] and H$\alpha$ (F502N and F656N, this column
contains the monochromatic line flux for the emission knot ($I_\lambda$ in $\rm
erg\,cm^{-2}\,s^{-1}$). }
\end{deluxetable}

\begin{deluxetable}{lccl}
\tablecaption{Astrometry of Compact Emission Knot Relative to Central Star} 
\tablewidth{0pt}
\tablehead{
\colhead{Date} &
\colhead{Separation [$''$]} &
\colhead{PA (J2000) [$^\circ$]} &
\colhead{Source} 
}
\startdata 
1991.9198 & $0.173\pm0.009 $ & $266.3\pm1.7$ & Liebert et al.\ (2013) \\
1993.0942 & $0.156\pm0.009 $ & $268.9\pm1.7$ & \qquad$''$ \\
1995.7768 & $0.162\pm0.009 $ & $269.6\pm1.7$ & \qquad$''$ \\
2013.9314 & $0.163\pm0.003 $ & $267.9\pm1.4$ & This paper\tablenotemark{a} \\
\enddata 
\tablenotetext{a}{Errors were estimated from the agreement between the [\OIII]
and H$\alpha$ measurements.}
\end{deluxetable}

\begin{deluxetable}{lccccl}
\tablecaption{Absolute {\em BVRI\/} Photometry of Central Star} 
\tablewidth{0pt}
\tablehead{
\colhead{UT Date}  &
\colhead{$V$}   &
\colhead{$B-V$} &
\colhead{$V-R$} &
\colhead{$V-I$} &
\colhead{Source} 
}
\startdata
1978 Apr 10 & 16.05  & $-0.24 $ & $\dots $ & $\dots $ & L89		\\
1982 Apr 24 & 16.00  & $-0.34 $ & $\dots $ & $\dots $ & EGB; L89	\\
1991 Jan 9  & 15.992 & $-0.315$ & $-0.136$ & $-0.273$ & This paper      \\
1994 May 20 & 15.995 & $-0.309$ & $-0.123$ & $-0.260$ & \qquad$''$      \\
1998 Mar 20 & 16.018 & $-0.329$ & $\dots $ & $-0.323$ & \qquad$''$      \\
2007 Nov 2  & 16.001 & $-0.310$ & $-0.128$ & $-0.291$ & De Marco et al.\ 2013 \\
2007 Nov 5  & 15.997 & $-0.303$ & $-0.148$ & $-0.314$ & \qquad$''$      \\    
2011 Mar 11 & 15.991 & $-0.303$ & $\dots $ & $-0.340$ & Douchin et al.\ 2015 \\
2011 Mar 16 & 16.004 & $-0.313$ & $\dots $ & $-0.319$ & \qquad$''$      \\
2016 Mar 14 & 15.992 & $-0.302$ & $-0.125$ & $-0.327$ & This paper      \\
2016 Mar 17 & 15.989 & $-0.306$ & $-0.145$ & $-0.346$ & \qquad$''$      \\
\noalign{\vskip0.125in} 
Mean\tablenotemark{a} \&     & 15.998 & $-0.310$ & $-0.134$ & $-0.310$ & \\
\quad error & $\pm$0.003 & $\pm$0.003 & $\pm$0.004 & $\pm$0.010 &       \\
\enddata 
\tablenotetext{a}{The photoelectric measurements from 1978 \& 1982 are not
included in the means}
\end{deluxetable}

\begin{deluxetable}{lcccccc}
\tablecaption{Equivalent Widths [\AA] for EGB 6 Emission Lines} 
\tablewidth{0pt}
\tablehead{
\colhead{Date}  &
\colhead{No.\ of}  &
\colhead{[\NeIII]}  &
\colhead{[\OIII]}  &
\colhead{H$\beta$}  &
\colhead{[\OIII]}  &
\colhead{[\OIII]}  \\
\colhead{Interval}  &
\colhead{Spectra}  &
\colhead{$\lambda$3869}  &
\colhead{$\lambda$4363}  &
\colhead{}  &
\colhead{$\lambda$4959}  &
\colhead{$\lambda$5007}
}  
\startdata
2004.1--2004.3 & 2 &     2.0 &    0.9 &    2.1 &    5.8 &    18.7 \\
2004.9--2005.4 & 6 &     2.3 &    1.0 &    2.3 &    5.5 &    16.8 \\
2005.9--2006.4 & 6 &     2.2 &    1.0 &    2.6 &    5.8 &    17.8 \\
2007.0--2007.3 & 5 &     2.1 &    0.9 &    2.2 &    5.5 &    17.2 \\
2007.9--2008.5 & 6 &     1.7 &    1.2 &    2.2 &    5.5 &    16.6 \\
2008.9--2009.3 & 7 &     2.0 &    1.2 &    2.4 &    5.5 &    17.3 \\
2010.0--2010.4 & 6 &     1.8 &    1.2 &    2.4 &    5.8 &    17.4 \\
2011.0--2011.2 & 5 &     1.8 &    1.2 &    2.6 &    5.9 &    17.9 \\
2011.9--2012.1 & 4 &     1.5 &    1.3 &    2.6 &    6.4 &    17.8 \\
\noalign{\vskip0.125in} 
Mean 1.5-m spectrum  & 47 &    1.9 &    1.1 &    2.4 &    5.7 &    17.4 \\
\noalign{\vskip0.125in} 
1978--1982\tablenotemark{a} & & 1.7 &    1.1 &    2.9 &    5.4 &    17.9 \\
\noalign{\vskip0.125in}
2016.25 Asiago 1.22-m\tablenotemark{b} & & 1.7 & 1.7 & 1.9 &    6.4 &    18.4 \\
\enddata 
\tablenotetext{a}{From L89, Table~2C; these are mean equivalent widths from
spectra obtained in 1978--1987}
\tablenotetext{b}{Asiago 1.22-m, 2016 April~3}
\end{deluxetable}

\begin{deluxetable}{lcccc}
\tablecaption{Spectral Energy Distribution of EGB~6 Central Source} 
\tablewidth{0pt}
\tabletypesize{\scriptsize}
\tablehead{
\colhead{Bandpass}  &
\colhead{Magnitude}  &
\colhead{Source\tablenotemark{a}}  &
\colhead{$\lambda_{\rm eff}$\tablenotemark{b}}  &
\colhead{$F_\lambda$\tablenotemark{c}}  \\
\colhead{ }  &
\colhead{ }  &
\colhead{ }  &
\colhead{[$\mu$m]}  &
\colhead{[$\rm erg\,cm^{-2}\,s^{-1}\,\AA^{-1}$]}  
}  
\startdata 
FUV   & 13.75  & GALEX    & 0.1516 & $1.50\times10^{-13}$  \\
NUV   & 14.58  & GALEX    & 0.2267 & $3.12\times10^{-14}$  \\
\noalign{\vskip0.1in} 
$U$   & 14.465 & D15	  & 0.3597 & $6.98\times10^{-15}$  \\
$B$   & 15.688 & Table 3  & 0.4386 & $3.43\times10^{-15}$  \\
$V$   & 15.998 & Table 3  & 0.5491 & $1.49\times10^{-15}$  \\
$R$   & 16.132 & Table 3  & 0.6500 & $7.79\times10^{-16}$  \\
$I$   & 16.308 & Table 3  & 0.7884 & $3.50\times10^{-16}$  \\
\noalign{\vskip0.1in} 
$u$   & 15.199 & D15 	  & 0.3586 & $7.31\times10^{-15}$  \\
$g$   & 15.660 & D15 	  & 0.4716 & $2.66\times10^{-15}$  \\
$r$   & 16.216 & D15 	  & 0.6165 & $9.34\times10^{-16}$  \\
$i$   & 16.563 & D15 	  & 0.7475 & $4.62\times10^{-16}$  \\
$z$   & 16.906 & D15 	  & 0.8922 & $2.32\times10^{-16}$  \\
\noalign{\vskip0.1in} 
$J$   & 16.38  & FL93	  & 1.237  & $8.70\times10^{-17}$  \\
$H$   & 16.08  & FL93	  & 1.645  & $4.18\times10^{-17}$  \\
$K$   & 15.65  & FL93	  & 2.212  & $2.20\times10^{-17}$  \\
\noalign{\vskip0.1in} 
$J$   & 16.518 & 2MASS    & 1.241  & $7.66\times10^{-17}$  \\
$H$   & 15.945 & 2MASS    & 1.651  & $4.72\times10^{-17}$  \\
$K_s$ & 16.099 & 2MASS    & 2.165  & $1.55\times10^{-17}$  \\
\noalign{\vskip0.1in} 
$W$1  & 14.651 & AllWISE  & 3.37   & $1.13\times10^{-17}$  \\
$W$2  & 13.636 & AllWISE  & 4.62   & $8.47\times10^{-18}$  \\
$W$3  & 9.520  & AllWISE  & 12.08  & $1.01\times10^{-17}$  \\
$W$4  & 7.170  & AllWISE  & 22.19  & $6.90\times10^{-18}$  \\
\noalign{\vskip0.1in} 
IRAC 1& $\dots$& C11      & 3.550  & $2.32\times10^{-17}$ \\
IRAC 2& $\dots$& C11	  & 4.493  & $1.75\times10^{-17}$ \\
IRAC 3& $\dots$& C11	  & 5.731  & $1.62\times10^{-17}$ \\
IRAC 4& $\dots$& C11	  & 7.872  & $1.82\times10^{-17}$ \\
MIPS 1& $\dots$& C11	  & 23.68  & $6.28\times10^{-18}$ \\
\noalign{\vskip0.1in} 
PACS  & $\dots$& Herschel & 70     & $2.20\times10^{-19}$ \\
\enddata 
\tablenotetext{a}{Sources for magnitudes or fluxes are: D15 (Douchin et al.\
2015, who quote values from SDSS); FL93 (Fulbright \& Liebert 1993, mean
values); the Mikulski Archive for Space Telescopes ({\tt
http://galex.stsci.edu}) for \GALEX; the NASA/IPAC Infrared Science Archive
({\tt http://irsa.ipac.caltech.edu/frontpage}) for 2MASS and \WISE; C11 (Chu et
al.\ 2011) for \Spitzer\/ IRAC and MIPS; our \S4.1 for {\it Herschel}.}
\tablenotetext{b}{Effective wavelengths are from {\tt http://galex.stsci.edu}
for \GALEX; {\tt http:/\slash www.astro.ucla.edu/$^\sim$wright\slash WISE\slash
passbands.html} for the \WISE\/ bands; {\tt
http://irsa.ipac.caltech.edu/data/SPITZER/docs\slash
irac/iracinstrumenthandbook/} and {\tt http://irsa.ipac.caltech.edu\slash
data\slash SPITZER/docs\slash mips/mipsinstrumenthandbook/} for \Spitzer\/ IRAC
and MIPS; and Douchin et al.\ (2015) for {\it UBVRI}, {\it ugriz}, {\it JHK},
and {\it JHK$_s$}.}
\tablenotetext{c}{Photometric zero-points are from {\tt http://galex.stsci.edu}
for \GALEX; Mann \& von Braun (2015) for {\it UBVRI\/} and ground-based {\it
JHK}; from the compilation at {\tt http://coolwiki.ipac.caltech.edu\slash
index.php\slash Central\_wavelengths\_and\_zero\_points} for 2MASS and AllWISE;
and from {\tt http://www.sdss.org/dr12/algorithms/fluxcal} for SDSS.}
\end{deluxetable}

\begin{figure}
\begin{center}
\plotone{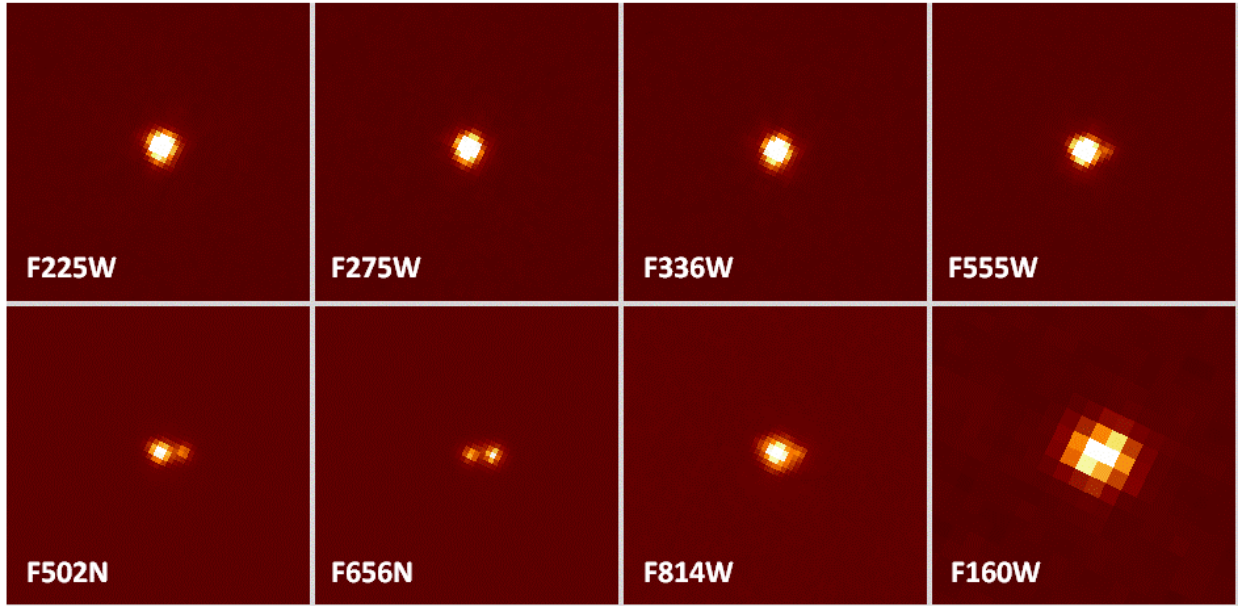}
\end{center}
\vskip-0.2in 
\figcaption{
\HST/WFC3 images of the nucleus of EGB~6, obtained in 2013 December. Each frame
is $2\farcs2\times2\farcs2$ and has north at the top and east on the left.
{\bf Top row:} Broad-band images in near-UV (F225W and F275W), $U$ (F336W), and
$V$ (F555W) filters. The near-UV and $U$-band images show only the hot central
white dwarf. The $V$-band image shows the companion emission knot faintly,
because [\OIII]~5007~\AA\ and other emission lines are included in the bandpass. 
{\bf Bottom row:} Narrow-band images in [\OIII]~5007~\AA\ (F502N) and H$\alpha$
(F656N), and broad-band images in $I$ (F814W) and $H$ (IR-channel F160W). The
narrow-band images separate the unresolved emission knot, lying $0\farcs163$ to
the west, from the central star. The $I$-band image shows a weak signal from the
photosphere of the cool dM companion, lying at the location of the compact
emission knot. In the $H$ band, the pair is not resolved due to the larger
pixels.
}
\end{figure}

\begin{figure}
\begin{center}
\plotone{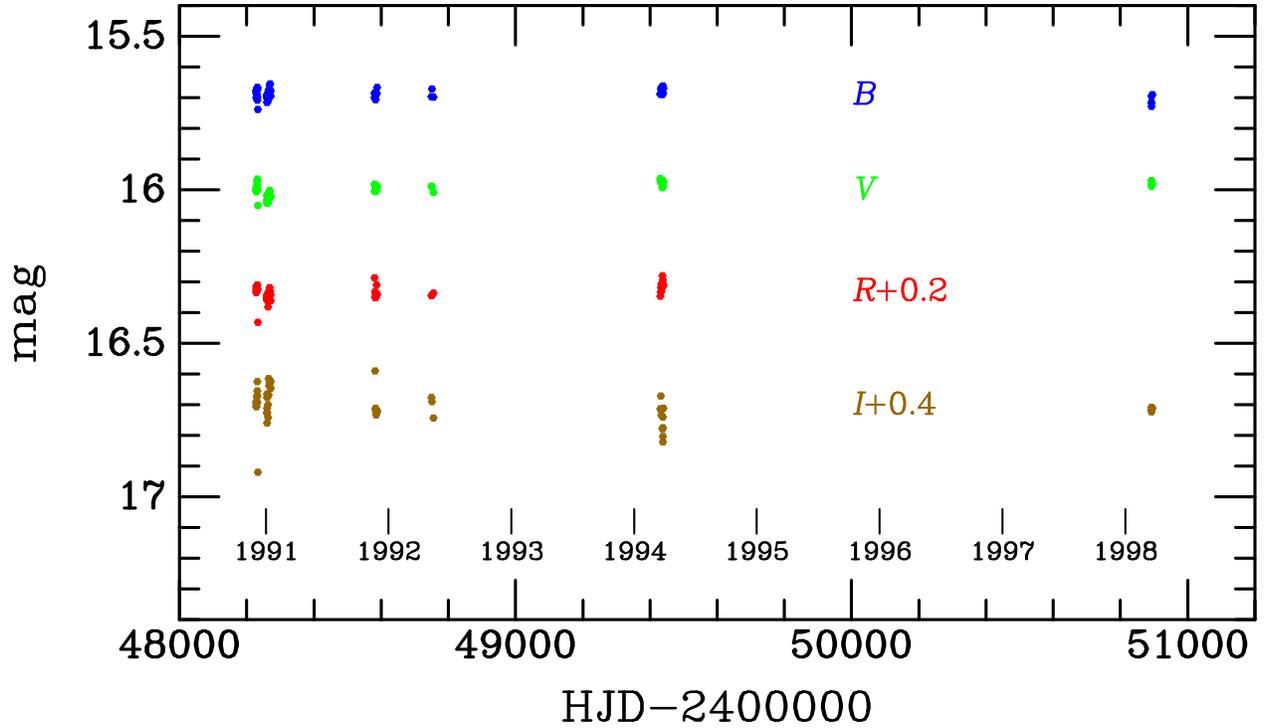}
\end{center}
\vskip-0.2in 
\figcaption{
Differential photometry of the central source in EGB~6, obtained with 0.9-m
telescopes at KPNO and CTIO in $B$, $V$, $R$, and $I$, plotted with blue, green,
red, and brown points, respectively. The zero-points of this
relative photometry have been adjusted to reproduce the mean absolute values in
Table~3, but $R$ and $I$ have been offset for clarity by the amounts indicated.
The $B$ and $V$ magnitudes appear to be constant within the uncertainties, but
there is a suggestion of variability in $R$ and more clearly in $I$.
}
\end{figure}

\begin{figure}
\begin{center}
\includegraphics[width=3.75in]{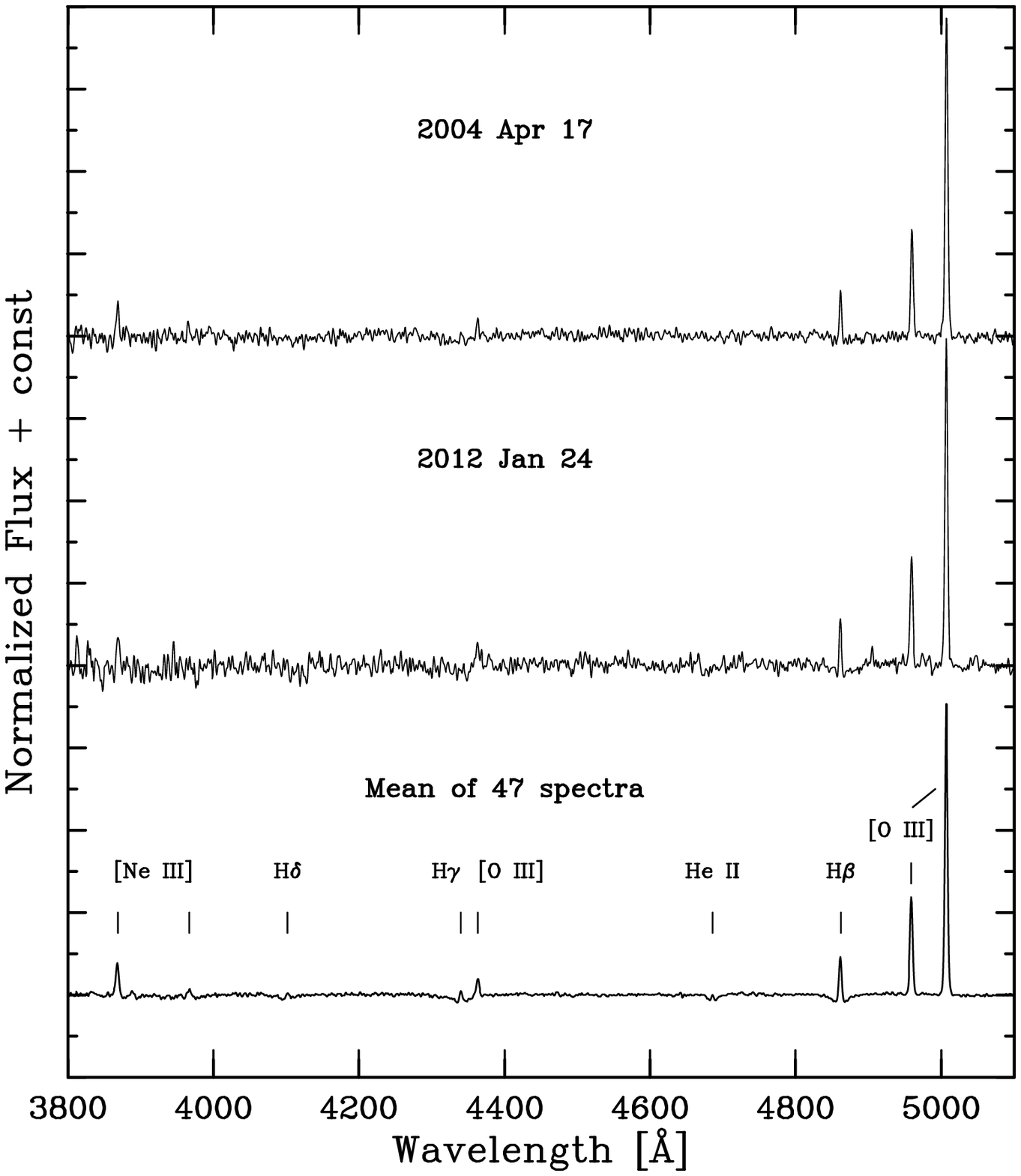}
\vskip0.1in
\includegraphics[width=3.75in]{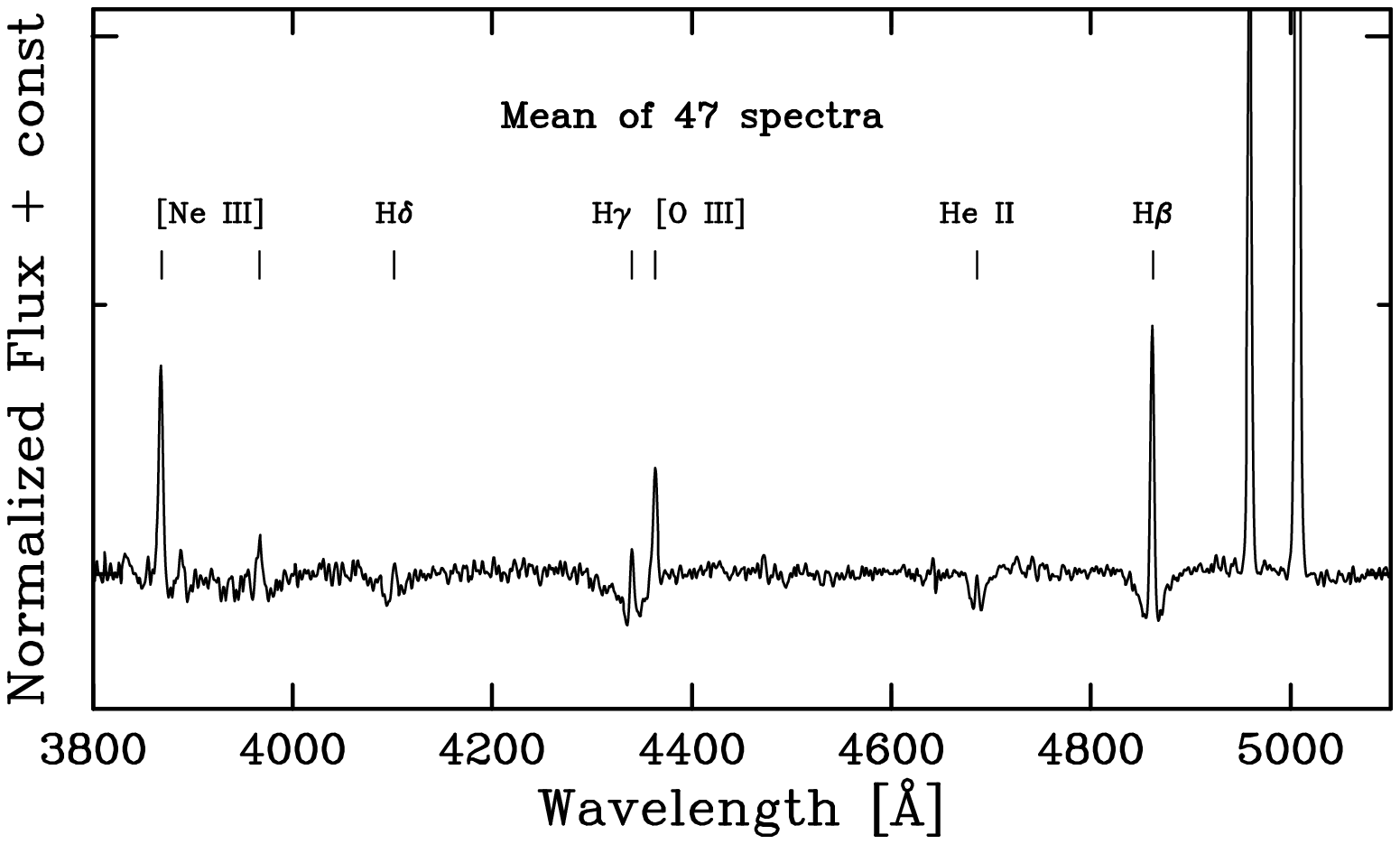}
\end{center}
\vskip-0.2in 
\figcaption{
{\it Top panel}: Spectra of the EGB~6 central source obtained with the SMARTS
1.5-m RC spectrograph (resolution $\sim$4.3~\AA), normalized to a flat
continuum. Two individual spectra (exposure time 1200~s) are shown to illustrate
their quality, along with the mean of all 47 spectra obtained between 2004 and
2012. We see no evidence for changes in the spectra during this interval.
{\it Bottom panel}: The mean spectrum with an expanded vertical scale, to show
photospheric absorption features more clearly. In both panels the tick
marks on the vertical scales are spaced at 0.5 of the continuum level.
}
\end{figure}

\begin{figure}
\begin{center}
\plotone{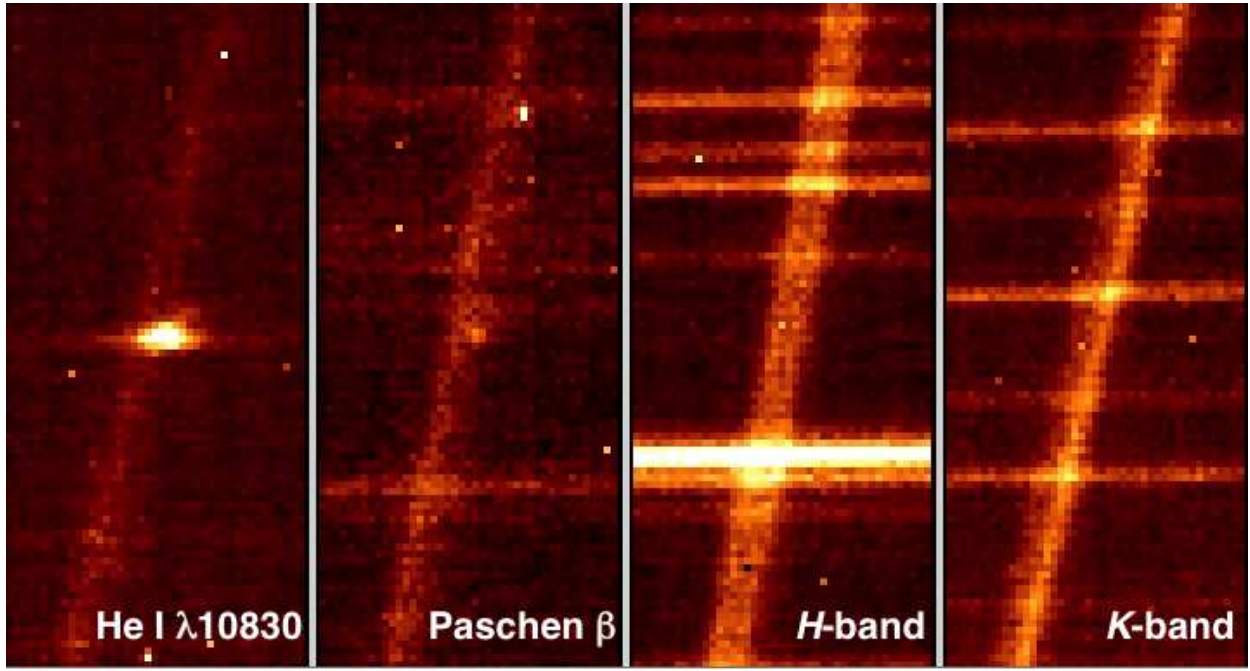}
\end{center}
\vskip-0.2in 
\figcaption{
False-color representations of four sections of the two-dimensional NIR Gemini
long-slit spectrum of EGB~6. The slit was oriented along the line joining the WD
and cool companion, with the WD lying on the left (east) side. The spatial width
of each panel is $2\farcs2$, the same as for the \HST\/ images in Fig.~1.
Emission lines extending across the entire slit length are mostly due to
terrestrial OH\null. The height of each panel in the dispersion direction is
$\sim$290~\AA\null. In the first two panels, the emission lines of \HeI\
10830~\AA\ and Paschen-$\beta$ are spatially offset from the spectrum of the WD,
consistent with the location of the emission knot seen in the narrow-band \HST\/
images. At the $H$ band in the third panel, the continua of the WD and cool
companion source have comparable fluxes. In the $K$ band shown in the fourth
panel, the cool companion's spectrum dominates over that of the WD.
}
\end{figure}

\begin{figure}
\begin{center}
\plotone{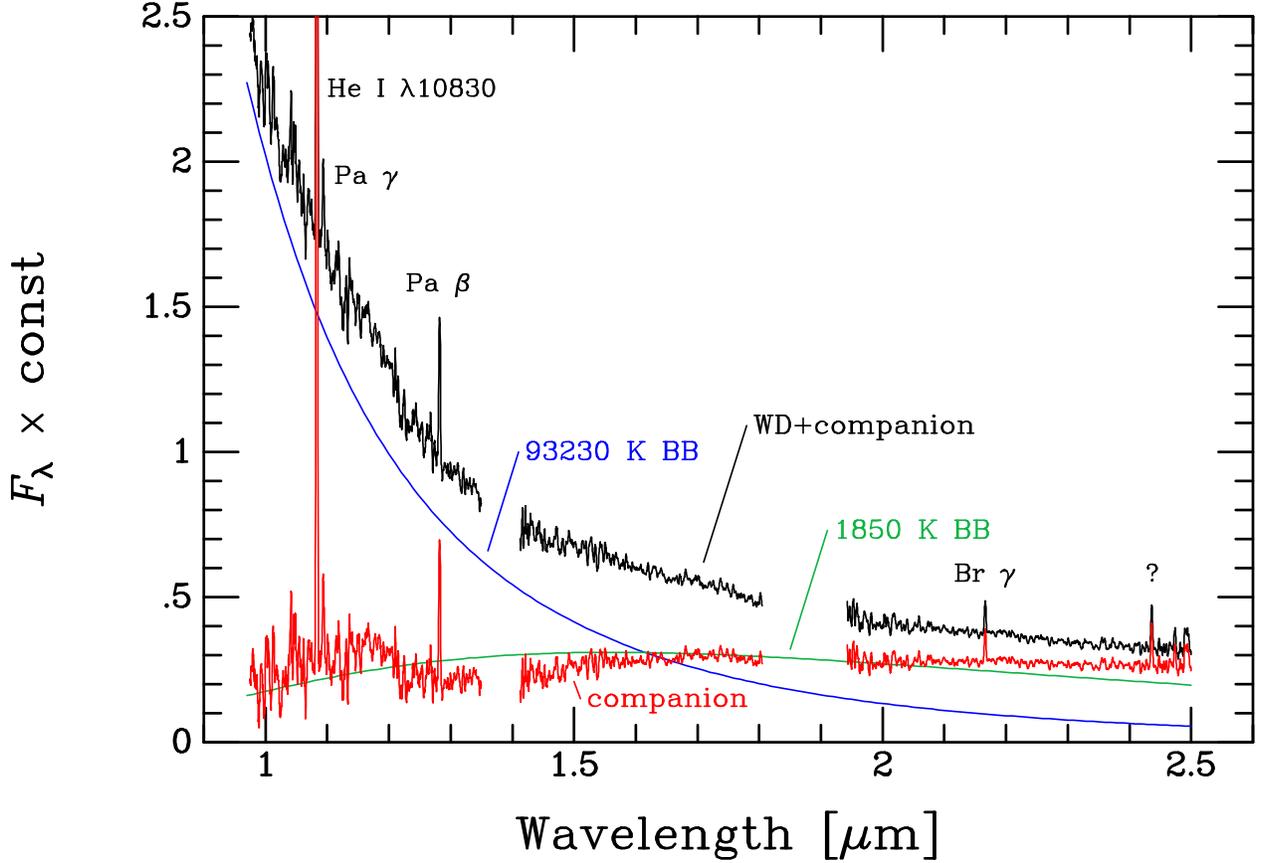}
\end{center}
\vskip-0.2in 
\figcaption{
The Gemini North NIR spectrum of the central source in EGB~6. The {\it black
line\/} plots the combined spectrum of the WD and the cool companion, which has
been smoothed with an 11-point boxcar kernel. Prominent emission lines from the
compact dense nebula are marked. To recover the spectrum of the companion, we
represent the WD spectrum as a 93,230~K blackbody, normalized to have a flux in
the center of the $H$ band equal to that of the companion, as shown in the third
panel of Fig.~4; this is plotted as a {\it blue curve}. Subtracting this
blackbody results in the companion spectrum, shown as a {\it red line}. Overall,
its energy distribution can be fitted approximately to a blackbody of about
1850~K ({\it green curve}). The bump from $\approx$1 to $1.2\,\mu$m is of
doubtful reality (see text).
}
\end{figure}

\begin{figure}
\begin{center}
\plotone{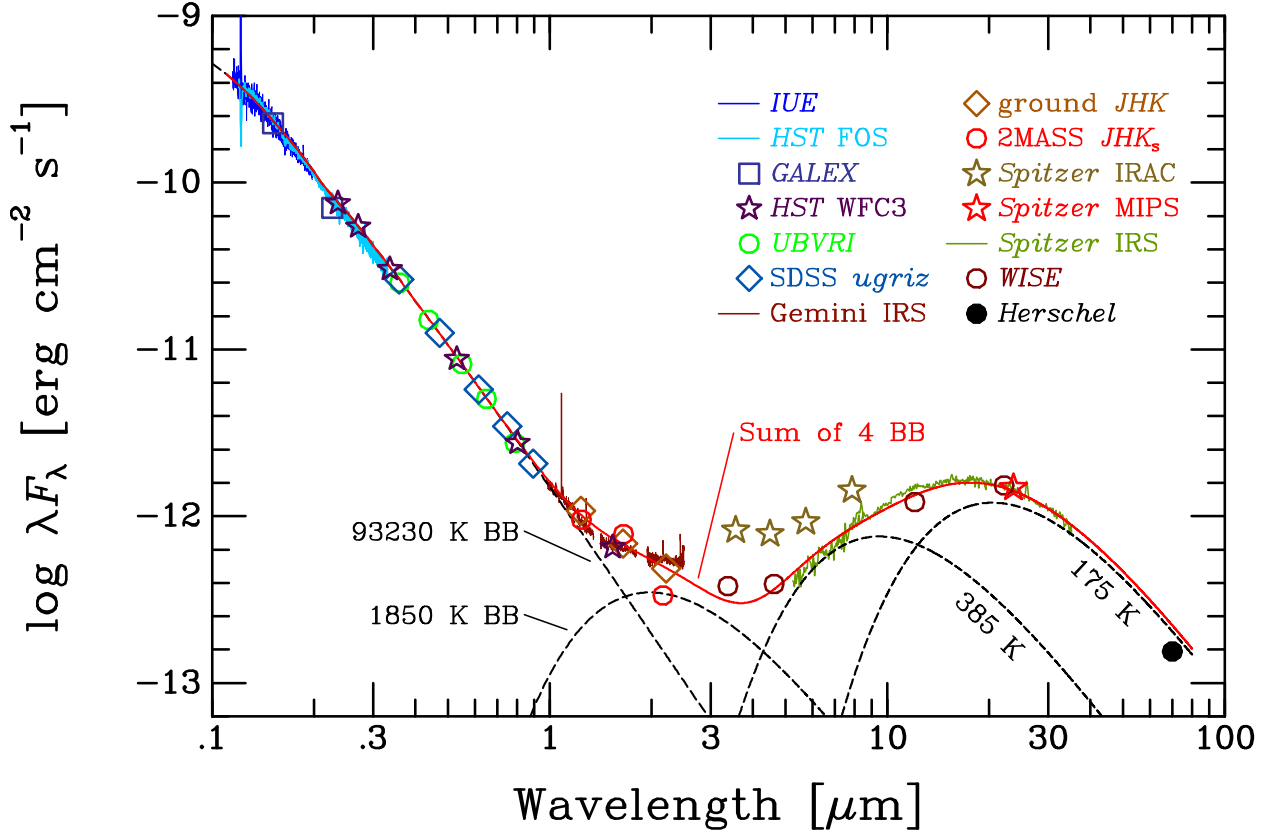}
\end{center}
\vskip-0.2in 
\figcaption{
Spectral-energy distribution for the EGB~6 nucleus. Observed photometric and
spectroscopic data are indicated by points and continuous lines, as indicated in
the legend within the figure. The strong emission line in the Gemini IRS
spectrum is \ion{He}{1} 10830~\AA\null. As described in the text, the SED has
been fitted by four blackbodies, represented by the {\it dashed black lines},
whose temperatures are indicated. In this fitting, the four bright \Spitzer\/
IRAC points were disregarded. A small reddening of $E(B-V)=0.02$ has been
applied. The {\it continuous red line\/} shows the sum of the fluxes of the four
blackbodies. Alternatively, the near- and mid-IR SED could be attributed to an
accretion disk with a range of temperatures (see text).
}
\end{figure}

\begin{figure}
\begin{center}
\includegraphics[width=5.5in]{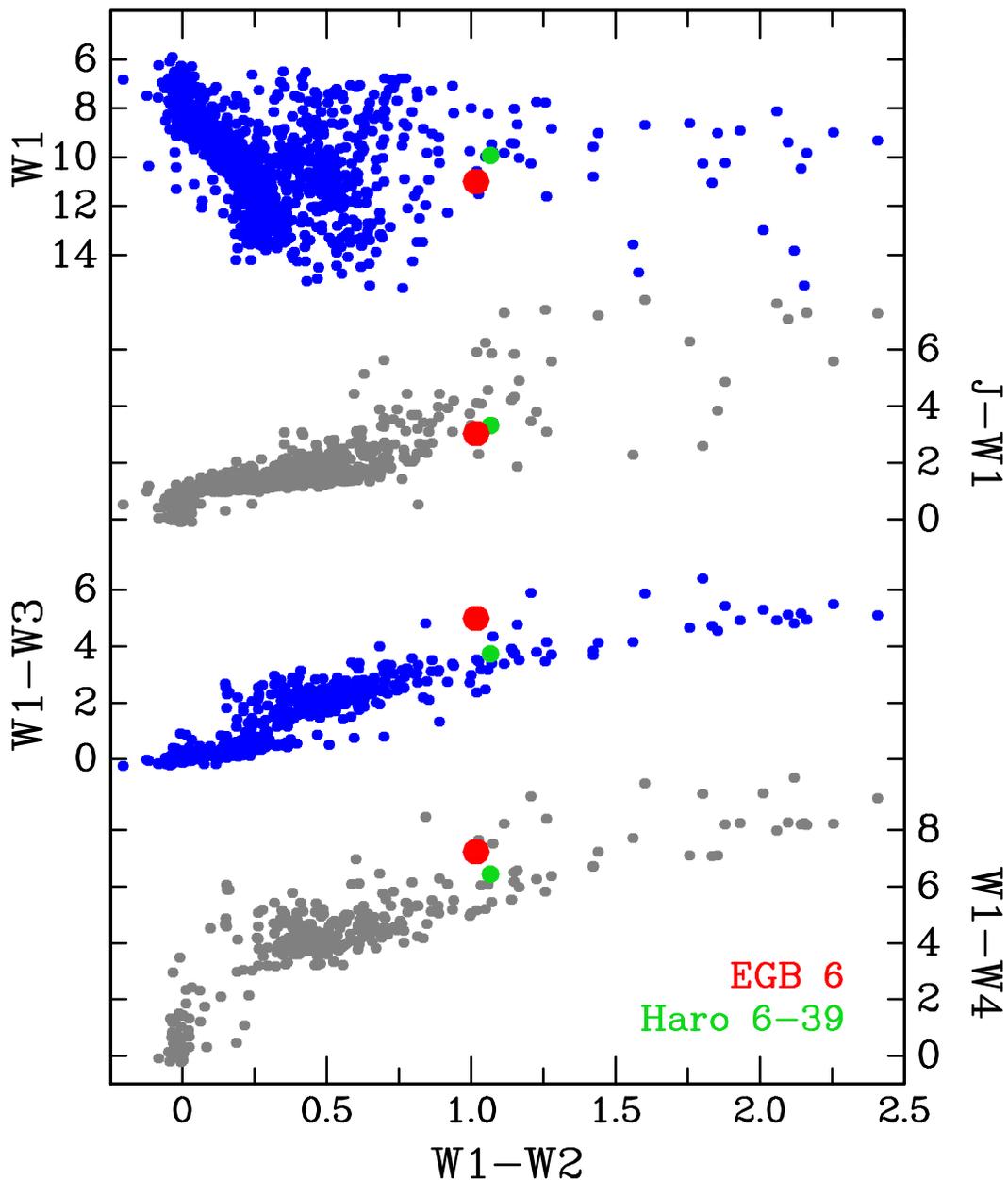}
\end{center}
\vskip-0.2in 
\figcaption{\small
NIR/MIR color-magnitude and color-color diagrams for samples of YSOs in the
Taurus and Upper Scorpius star-forming regions (blue and grey points), taken
from the 2MASS and \WISE\/ all-sky surveys (see text for details and
references). Upper Scorpius $W1$ magnitudes have been adjusted to the distance
of Taurus. Shown as {\it filled red circles\/} are the data for the EGB~6
companion (with the contribution from the hot WD subtracted, and the $W1$
magnitude adjusted from 725~pc to the 140~pc distance of the Taurus region). The
{\it filled green circles\/} highlight the Taurus YSO Haro~6-39, chosen because
its colors are similar to those of EGB~6; moreover, its NIR spectrum resembles
that of the EGB~6 companion.
}
\end{figure}

\end{document}